\documentclass{article}
\title{Continuous and robust clustering coefficients \\ for weighted and directed networks}
\usepackage{graphics}
\usepackage{graphicx}
\usepackage{epsfig}
\usepackage{amssymb}
\usepackage{amsthm}
\usepackage{amsmath,amssymb}
\usepackage{amsbsy}
\usepackage{multirow}

\makeatletter
\renewcommand{\section}{%
  \@startsection{section}%
   {1}%
   {\z@}%
   {-3.5ex \@plus -1ex \@minus -.2ex}%
   {2.3ex \@plus.2ex}%
   {\normalfont\normalsize\bfseries}%
}%
\renewcommand{\subsection}{%
  \@startsection{subsection}%
   {1}%
   {\z@}%
   {-3.5ex \@plus -1ex \@minus -.2ex}%
   {2.3ex \@plus.2ex}%
   {\normalfont\small\bfseries}%
}%
\renewcommand{\subsubsection}{%
  \@startsection{subsubsection}%
   {1}%
   {\z@}%
   {-3.5ex \@plus -1ex \@minus -.2ex}%
   {2.3ex \@plus.2ex}%
   {\normalfont\small\bfseries}%
}%
\makeatother

\author{
    Kent Miyajima (miyajima@i.h.kyoto-u.ac.jp)\\
  \and
    Takashi Sakuragawa (sakura@i.h.kyoto-u.ac.jp)\\
    {\small Graduate School of Human and Environmental Studies}, \\
    {\small Kyoto University, Yoshida, Sakyo-ku, Kyoto 606-8501, Japan}
}

\date{\today}

\begin{document}
\maketitle

\begin{abstract}
We introduce new clustering coefficients for weighted networks. They are continuous and robust against edge weight changes.
Recently, generalized clustering coefficients for weighted and directed networks have been proposed. 
These generalizations have a common property, that their values are not continuous. They are sensitive with edge weight changes, especially at zero weight.
With these generalizations, if vanishingly low weights of edges are truncated to weight zero for some reason, the coefficient value may change significantly from the original value. It is preferable that small changes of edge weights cause small changes of coefficient value. We call this property the \textit{continuity} of generalized clustering coefficients. 
Our new coefficients admit this property. In the past, few studies have focused on the continuity of generalized clustering coefficients. 
In experiments, we performed comparative assessments of existing and our generalizations. In the case of a real world network dataset (C. Elegans Neural network), after adding random edge weight errors, though the value of one discontinuous generalization was changed about 436 \%, the value of proposed one was only changed 0.2 \%.

\end{abstract}

\section{Introduction}
\label{section1}

In this paper, we propose generalized clustering coefficients for weighted and directed networks. 

The clustering coefficient was introduced by Watts and Strogatz to analyze unweighted and undirected networks \cite{watts1998collective}. 
It is a measure of the likelihood that two nodes connecting to a certain node are connected to each other. 
For example, in social network study, the clustering coefficient is used to measure the probability that a friend of a person is also his/her friend. 
It is a common useful tool for analyzing real world networks.

The original clustering coefficient cannot be applied to weighted or directed networks. 
However, there are a lot of weighted or directed real world networks (e.g., C.elegans' neural network \cite{watts1998collective}, scientists' collaboration network \cite{barrat2004architecture}). 
To calculate the clustering coefficients of this kind of network, we need to convert them to unweighted and undirected networks by normalizing edge weights to 0 or 1 and treating directed edges as undirected edges. Such conversion may distorse the information contained in the network.
Therefore, a number of researchers have proposed generalized clustering coefficients for weighted or directed networks which can be directly calculated from the original network (e.g.,\cite{lopez2004applying,barrat2004architecture,onnela2005intensity}). 

In the past, generalized clustering coefficients for weighted and directed networks has been proposed by Fagiolo \cite{fagiolo2007clustering}, Suzuki \cite{suzuki2009fukuzatsu} and Opsahl et al. \cite{opsahl2009clustering}. 
The advantage of these generalizations is that these deal directly with weighted and directed edges without conversion.

However, these generalizations for weighted and directed networks have a common property, that the value of proposed clustering coefficient is not continuous. They are sensitive with changes of edge weights, especially at zero weight.
With these coefficients, if vanishingly low weights of edges are truncated to weight zero for some reason, the coefficient value may change significantly from the original value. It is preferable that small changes of edge weights cause small changes of the coefficient value. We call this property the \textit{continuity} of generalized clustering coefficients. 

In this paper, we propose continuous generalizations of the clustering coefficient. Few studies have focused on the continuity of generalized clustering coefficients. 
A continuous clustering coefficient reduces the range of the error associated with edge weight errors.

The remainder of this paper is as follows. 
In section \ref{section2} we briefly introduce the original clustering coefficient and list its existing generalizations. Then we summarize their properties. We also state about the measures of the continuity.
In section \ref{section3} we propose generalized clustering coefficients for weighted and directed networks, and summarize their properties.
In section \ref{section4} we conduct numerical experiments and show that our continuous clustering coefficients are robust against edge weight errors.
In section \ref{section5} we discuss the results of the experiments.

\section{Existing generalizations of clustering coefficient}
\label{section2}

In this section, first we briefly introduce the clustering coefficient \cite{watts1998collective,newman2002random}. 
Second we list its existing generalizations for weighted and undirected networks \cite{latora2003economic,lopez2004applying,barrat2004architecture,onnela2005intensity,zhang2005general,kalna2006clustering,holme2007korean,abdallah2009new}. Then we list the generalizations for weighted and directed networks \cite{fagiolo2007clustering,suzuki2009fukuzatsu,opsahl2009clustering}. 
At the end of this section, we summarize their properties.

In this paper we use the following notations. Let $G = (V,\:\{w_{ij}\}_{i,\:j \: \in \: V})$ denote a simple graph\footnote{A graph that has no multiple edge and no self-loop. We use the words \textit{graph} and \textit{network} interchangeably. } with the set of nodes $V \subset \mathbb{N}$ and the weights $w_{ij} \in \{0,1\}$ (or $[0,1]$) between node $i$ and node $j$. Note that the weight zero means there is no edge. Let $N$ denote the number of nodes in the network. 

In an undirected network, $w_{ij}$ are symmetric, i.e. $w_{ij} = w_{ji}$. Let $V_k$ denote the set of the neighbors of node $k$ (i.e., $V_k = \{i \in V \;|\; w_{ki} >0\}$). Let $N_k$ denote the cardinality of $V_k$.  $N_k$ is called the degree of node $k$.
Let $T_k$ denote the number of the triangles of node $k$, which is defined as the element count of $\{ (i,j) \in V \times V \;|\; i < j ,\; w_{ik} > 0 ,\; w_{jk} > 0 \;\textrm{and}\; w_{ij} > 0 $ \}.

In a directed network, we use $w_{i \rightarrow j}$ instead of $w_{ij}$ and they are not necessarily symmetric, i.e., $w_{i \rightarrow j}$ denote the weight of an edge from node $i$ to node $j$. 

In an unweighted network, $w_{ij} \in \{0,1\}$ and $w_{i \rightarrow j} \in \{0,1\}$. In a weighted network, $w_{ij} \in [0,1]$ and $w_{i \rightarrow j} \in [0,1]$.

\subsection{Clustering coefficients}\label{clusteringcoefficient}

The clustering coefficient is a measure of the tendency to form tightly connected neighborhoods. 
The clustering coefficient of node $k$ with $N_k \geq 2$ is defined as follows:
\begin{eqnarray}\label{watts}
C_{ws}(k) = \frac{T_k}{{\scriptscriptstyle{}_{N_k}} \textrm{C}_2 \;\;} = \frac{2\;T_k}{N_k(N_k-1)}. 
\end{eqnarray}
Note that $C_{ws}(k)$ is undefined when $N_k \leq 1$.\footnote{For practical use, in the case that $N_k \leq 1$ some researchers defined $C(k) = 0$ or defined them by means of an indeterminate \cite{ide2007limit,konno_ide200805}.}

The clustering coefficient of the whole network is defined as the average over the clustering coefficients of all nodes as follows: 
\begin{eqnarray}\label{wattslocal}
C_{loc} = \frac{1}{N} \;\displaystyle\sum_k C_{ws}(k). 
\end{eqnarray}
Note that $C_{loc}$ is undefined when there is any node whose coefficient is undefined. $C_{ws}(k),\; C_{loc} \in [0,1]$ because each numerator is a part of the corresponding denominator in Eq.(\ref{watts}) and Eq.(\ref{wattslocal}) respectively.

Another definition of clustering coefficient was proposed by Newman, Watts and Strogatz \cite{newman2002random} as follows: 
\begin{eqnarray}\label{newmanglobal}
C_{glo} = \frac{3 \times \textrm{number of triangles in the graph}}{\textrm{number of connected triples of nodes}}. 
\end{eqnarray}
Here \textit{connected triples} are the trios of nodes that at least one node is connected to both of the other nodes. Note that $C_{glo}$ is undefined when all nodes have less than degree two. 
The clustering coefficient with this definition is sometimes called \textit{transitivity} (\cite{schank2005approximating}). 
Usually, there is a difference between the values of (\ref{wattslocal}) and (\ref{newmanglobal}). 
In this paper, we call the former the \textit{local clustering coefficient} and the latter the \textit{global clustering coefficient}\footnote{Some researchers refer $C_{ws}(k)$ as the local clustering coefficient and $C_{loc}$ as the global clustering coefficient \cite{ide2007limit}.}.

The clustering coefficient is defined without taking into consideration the edge weights and directions in the network. 
Thus we need to treat all edges as unweighted and undirected when we calculate the coefficient. 
In consequence of that conversion, the same result might be attributed to the networks that share the same topology but differ in the weights or directionality. For this reason, a number of researchers have proposed the generalizations of the local and global clustering coefficients for weighted or directed networks.

\subsection{Generalized clustering coefficients}\label{2.2}

In this subsection, first we summarize the existing generalizations of the clustering coefficient for weighted and undirected networks. 
Then we summarize the generalizations for weighted and directed networks.

\subsubsection{Generalizations for weighted and undirected networks}

Here, we summarize the generalized clustering coefficients for weighted and undirected networks. 

\subsubsection*{Latora et al.}\label{latora1}

Latora et al. proposed the \textit{local efficiency} as a generalization of the global clustering coefficient as follows \cite{latora2003economic}:
\begin{eqnarray}\label{latora}
E_{loc} = \frac{1}{N} \displaystyle \sum_k \frac{E(\mathrm{G_k})}{E(\mathrm{G_k^{ideal}})}. 
\end{eqnarray}
Here $\mathrm{G_k}$ is the subgraph of the whole graph which consisted of node $k$ and its neighbors. $\mathrm{G_k^{ideal}}$ is the graph that has the same nodes of $\mathrm{G_k}$ but has all the $\frac{N_k(N_k - 1)}{2}$ possible edges. 
$E(\mathrm{G})$ is the average \textit{efficiency} of graph $\mathrm{G}$. It is defined as follows:
\begin{eqnarray}
E(\mathrm{G}) = \frac{\displaystyle \sum_{i,j} \; \epsilon_{ij}} {N (N - 1)} = \frac{1}{N (N - 1)} \displaystyle \sum_{i,j} \; \frac{1}{d_{ij}}. 
\end{eqnarray}
Here $d_{ij}$ is the shortest path length\footnote{The minimum number of the edges traversed to get from one node to the other node.} between node $i$ and node $j$, and $\epsilon_{ij}$ is defined as $\frac{1}{d_{ij}}$. 
Note that $d_{ij} = \infty$ and $\epsilon_{ij} = 0$ when there is no path between node $i$ and node $j$. 

\subsubsection*{Fernandez et al.}

Fernandez et al. proposed a generalization of the local clustering coefficient as follows \cite{lopez2004applying}:
\begin{eqnarray}\label{fernandez}
C_{fe}(k) = \displaystyle \sum_{i,j} \frac{w_{ij}}{ {}_{N_k} \mathrm{P}_2 }. 
\end{eqnarray}
Fernandez et al. stated that this equation can be interpreted as a measurement of the local efficiency (Eq.(\ref{latora})) of the network around node $k$.

\subsubsection*{Barrat et al.}

Barrat et al. proposed a generalization of the local clustering coefficient as follows \cite{barrat2004architecture}:
\begin{eqnarray}
C_{ba}(k) = \frac{1}{s_k(N_k - 1)} \displaystyle \sum_{i,j} \frac{(w_{ki} + w_{kj})}{2}\; a_{ik}a_{kj}a_{ij}. 
\end{eqnarray}
Here $s_k$ is defined as $\sum_j a_{kj}w_{kj}$ and $(a_{ij})$ is the adjacency matrix of the graph.

\subsubsection*{Onnela et al.}

Onnela et al. proposed a generalization of the local clustering coefficient as follows \cite{onnela2005intensity}:
\begin{eqnarray}\label{onnela}
C_{on}(k) = \frac{ \: \displaystyle \sum_{i,j} {(w_{ik}w_{kj}w_{ij})}^{\frac{1}{3}}} {N_k (N_k - 1)}. 
\end{eqnarray}

\subsubsection*{Zhang et al.}\label{zhang1}
Zhang et al. proposed a generalization of the local clustering coefficient as follows \cite{zhang2005general}:
\begin{eqnarray}\label{zhang}
C_{zh}(k) = \frac{\displaystyle \sum_{i,j} \; w_{ki}w_{ij}w_{jk}}{(\displaystyle \sum_i w_{ki})^2 - \sum_i w_{ki}^2}. 
\end{eqnarray}
This formula can be rewritten as follows \cite{kalna2006clustering,ahnert2007ensemble}:
\begin{eqnarray}\label{kalna}
C_{zh}(k) = \frac{\displaystyle \sum_{i,j} \; w_{ki}w_{kj}w_{ij}} {\displaystyle \sum_{i,j} \; w_{ki}w_{kj}}. 
\end{eqnarray}

\subsubsection*{Holme et al.}
Holme et al. proposed a generalization of the local clustering coefficient as follows \cite{holme2007korean}:
\begin{eqnarray}\label{holme}
C_{ho}(k) = \frac{\displaystyle \sum_{i,j}\;w_{ki}w_{ij}w_{jk}}{\mathrm{max}_{i,j}(w_{ij}) \; \displaystyle \sum_{i,j} \; w_{ki}w_{jk}}. 
\end{eqnarray}
Here $\mathrm{max}_{i,j}(w_{ij})$ is the maximum value of the weights in the network.

\subsubsection*{Opsahl et al.}\label{opsahl1}
\begin{figure}[htbp]
 \begin{center}
  \includegraphics[width=7.5cm,bb=0 0 468 208]{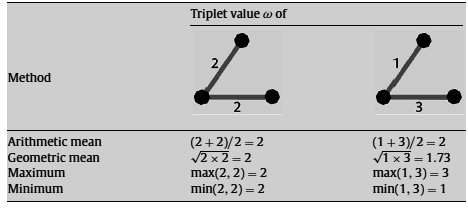}
 \end{center}
 \caption{Methods for calculating $\omega$ for each triplet (cited from Opsahl et al.  \cite{opsahl2009clustering}).}
 \label{fig:opsahl}
\end{figure}

Opsahl et al. proposed generalizations of the global clustering coefficient as follows \cite{opsahl2009clustering}:
\begin{eqnarray}\label{opsahl}
C_{op} = \frac{\rm{total\:value\:of\:closed\:triplets}}{\rm{total\:value\:of\:triplets}} = \frac{\displaystyle \sum_{\tau_\Delta} \omega}{\displaystyle \sum_{\tau} \omega}. 
\end{eqnarray}
Here $\tau$ is the set of triplets and $\tau_\Delta$ is the set of closed triplets. A triplet is defined as three nodes that are connected with either two (open triplet) or three (closed triplet) edges. $\omega$ is defined for each triplet. 
Opsahl et al. proposed four methods (arithmetic mean, geometric mean, maximum, minimum) for calculating the value of $\omega$ (Figure \ref{fig:opsahl}). 

\subsubsection*{Abdallah}
Abdallah proposed a generalization of the local clustering coefficient as follows\footnote{Abdallah stated that this generalization can also be defined for directed networks. But the detailed definition is not mentioned \cite{abdallah2009new}.} \cite{abdallah2009new}:
\begin{eqnarray}
C_{ab}(k) = \displaystyle \frac{c(E_k)}{ {}_{N_k} \mathrm{C}_2 } . 
\end{eqnarray}
Here $E_k$ is the set of edges between the neighbors of node $k$, and $c(E_k)$, \textit{effective cardinality}, is defined as follows:
\begin{eqnarray}
c(E_k) = 
\begin{cases}
0 &\textrm{if } E_k \textrm{ is empty},\\
2^\alpha & \textrm{otherwise}.
\end{cases}
\end{eqnarray}
where
\begin{eqnarray}
\alpha = \left(\displaystyle \sum_{e \in E_k} \frac{w(e)}{\displaystyle\sum_{o \in E_k} w(o)} \log_2 \frac{\displaystyle\sum_{o \in E_k} w(o)}{w(e)}\right).
\end{eqnarray}
$w(e)$ is the weight of edge $e \in E_k$. The value $\frac{w(e)}{\sum_{o \in E_k} w(o)}$ represents the probability of interaction over edge $e$ among all the edges in $E_k$.

\subsubsection{Generalizations for Weighted and Directed Networks}\label{yuukouomomituki}

Here, we summarize the generalized clustering coefficients for weighted and directed networks. 

\subsubsection*{Fagiolo}\label{fagiolo1}
Fagiolo proposed a generalization of the local clustering coefficient as follows\footnote{Fagiolo also proposed a generalization of the local clustering coefficient for unweighted and directed networks \cite{fagiolo2007clustering}.} \cite{fagiolo2007clustering}:
\begin{eqnarray}\label{fagiolo}
C_{fag}(k) = \displaystyle \sum_{i,j} 
{\textstyle 
\frac { (w_{k \rightarrow i}^{\frac{1}{3}} + w_{i \rightarrow k}^{\frac{1}{3}}) (w_{k \rightarrow j}^{\frac{1}{3}} + w_{j \rightarrow k}^{\frac{1}{3}}) (w_{i \rightarrow j}^{\frac{1}{3}} + w_{j \rightarrow i}^{\frac{1}{3}})} {2 \left[ d_k^{total} (d_k^{total} - 1) - 2 d_k^\leftrightarrow \right] }. 
}
\end{eqnarray}
Here $d_k^{total} = d_k^{in} + d_k^{out}$ is the total-degree of node $k$ and $d_k^\leftrightarrow = \sum_{j} a_{kj}a_{jk}$ is the number of the bilateral edges between node $k$ and its neighbors. 
$d_k^{in} = \sum_{j} a_{jk}$ is the in-degree of node $k$ and $d_k^{out} = \sum_{j} a_{kj}$ is the out-degree of node $k$.

\subsubsection*{Suzuki}\label{suzuki1}
Suzuki proposed a generalization of the local clustering coefficient as follows \cite{suzuki2009fukuzatsu}:
\begin{eqnarray}\label{suzuki}
C_{suz}(k) = \frac{1}{ {}_{N_k} \mathrm{P}_2 } \displaystyle \sum_{i,j} \frac { w_{k \rightarrow i} \: w_{i \rightarrow j} +  w_{k \rightarrow j} \: w_{j \rightarrow i} } {w_{k \rightarrow i} + w_{k \rightarrow j}}. 
\end{eqnarray}

\subsubsection*{Opsahl et al.}\label{opsahl2}
Opsahl et al. proposed generalizations of the global clustering coefficient not only to weighted and undirected networks but also weighted and directed networks \cite{opsahl2009clustering}. 

Also in the definition of the generalizations to weighted and directed networks, 
the same definition and methods are used as that in Eq.(\ref{opsahl}). 
However, unlike in the case of weighted and undirected networks, vacuous triplets \cite{wasserman1994social} are not part of the numerator nor of the denominator in Eq.(\ref{opsahl}). And the triplets which are non-transitive and non-vacuous are not part of the numerator. 

\subsection{Properties of Generalized Clustering Coefficients}\label{seisitu}

In this subsection, first we define properties of generalized clustering coefficients. Then we summarize the properties of the generalized clustering coefficients described in Section \ref{2.2}. At the end of this subsection, we briefly state about the measure of the continuity.

\subsubsection{Definitions of Properties}
Here, we define properties of generalized clustering coefficients.

\subsubsection*{Continuity}\label{renzoku}

The \textit{continuity} of a generalized clustering coefficient is defined as the property that small changes of edge weights cause only small changes of its value. Almost all coefficients in the last section are discontinuous and they may change significantly from the original value, if vanishingly low weights of edges are truncated to weight zero for some reason.

\subsubsection*{General versatility}\label{kakutyou}

\textit{General versatility} of a generalized clustering coefficient is defined as the property that its value coinsides with the value of the local or global clustering coefficient when edges are undirected (symmetric) and their weights are 0 or 1.

\subsubsection*{Weight-Scale Invariance}\label{huhen}

\textit{Weight-scale invariance} of a generalized clustering coefficient is defined as the property that its value is invariant when the weights of all edges in the network are multiplied by a positive constant.

When one experiments on a network dataset that edge weights are not normalized between 0 and 1, if one uses generalized clustering coefficients that have no weight-scale invariance, the resulting values vary significantly from the values that are calculated for the normalized dataset. This means without a normalization process, the selection of measurement scale affects significantly the coefficient value. When one uses generalized coefficients that have weight-scale invariance, one need not normalize edge weights in the network.

\subsubsection{Summaries of Properties}

Here, we summarize the properties of the generalized clustering coefficients in section \ref{2.2}. Table \ref{wuntable} and Table \ref{wdntable} shows the properties.

All of the existing generalized clustering coefficients for weighted and directed networks are discontinuous. In addition, there is no continuous generalization of the global clustering coefficient among them. In the next section, we propose continuous generalizations of the local and global clustering coefficients for weighted and directed networks.

\begin{table}[htbp]
\begin{center}
\caption{Properties of the existing generalizations for weighted and undirected networks.}
\label{wuntable}
\begingroup
\renewcommand{\arraystretch}{1.5}
\footnotesize
\begin{tabular}{|l|c|c|c|} \hline
             & Continuity & General Versatility & \shortstack{Weight-Scale \\ Invariance} \rule[0mm]{0mm}{6mm} \\\hline
$C_{ba}(k)$ &   & X &   \\\hline
$C_{ba}(k)$ &   & X & X \\\hline
$C_{on}(k)$ &   & X &   \\\hline
$C_{zh}(k)$ & X & X &   \\\hline
$C_{ho}(k)$ & X & X & X \\\hline
$C_{op}$ &   & X & X \\\hline
$C_{ab}(k)$ &   & X & X \\\hline
\end{tabular}
\endgroup
\end{center}
\end{table}
\begin{table}[htbp]
\begin{center}
\caption{Properties of the existing generalizations for weighted and directed networks.}
\label{wdntable}
\begingroup
\renewcommand{\arraystretch}{1.5}
\footnotesize
\begin{tabular}{|l|c|c|c|} \hline
             & Continuity & General Versatility & \shortstack{Weight-Scale \\ Invariance} \rule[0mm]{0mm}{6mm} \\\hline
$C_{fag}(k)$ &   & X &   \\\hline
$C_{suz}(k)$ &   & X &   \\\hline
$C_{ops}$ &   & X & X \\\hline
\end{tabular}
\endgroup
\end{center}
\end{table}

\subsubsection{Measures of Continuity}

Here, first we briefly introduce the measures of the continuity. Then we describe the continuity of the generalized clustering coefficients.

\subsubsection*{H\"{o}lder Continuity}

A function $f: X \rightarrow \mathcal{R}$ is \textit{H\"{o}lder continuous with exponent $\alpha$ $(0 < \alpha \leqq 1)$ on} $D \subset X$ if
\begin{eqnarray}\label{18}
\sup_{\substack{x,y \in D \\ x \neq y}} \frac{|f(x) - f(y)|}{|x - y|^\alpha} < +\infty.
\end{eqnarray}
And $f$ is \textit{H\"{older} continuous with exponent $\alpha$ at} $x_0 \in X$ if it is H\"{o}lder continuous on some neighborhood $D$ of $x_0$. 
\begin{eqnarray}\label{19}
\sup_{x \in D - \{x_0\} } \frac{|f(x) - f(x_0)|}{|x - x_0|^\alpha} < +\infty.
\end{eqnarray}
A function $f$ is \textit{locally H\"{o}lder continuous with exponent $\alpha$ on} $D \subset X$ if it is H\"{o}lder continuous with exponent $\alpha$ at every point of $D$.

\subsubsection*{Lipschitz Continuity}

When the sup in Eq.(\ref{18}) or Eq.(\ref{19}) is finite with exponent $\alpha =1$, $f$ is said to be \textit{Lipschitz continuous}. A function $f: X \rightarrow \mathcal{R}$ is \textit{Lipschitz continuous on} $D \subset X$ if
\begin{eqnarray}
\sup_{\substack{x,y \in D \\ x \neq y}} \frac{|f(x) - f(y)|}{|x - y|} < +\infty.
\end{eqnarray}
$f$ is \textit{Lipschitz continuous at} $x_0 \in X$ if it is Lipschitz continuous on some neighborhood $D$ of $x_0$. 
\begin{eqnarray}
\sup_{x \in D - \{x_0\} } \frac{|f(x) - f(x_0)|}{|x - x_0|} < +\infty.
\end{eqnarray}
A function $f$ is \textit{locally Lipschitz continuous on} $D \subset X$ if it is Lipschitz continuous at every point of $D$.

H\"{older} and Lipschitz continuities prove to be quantitative measures of the continuity. These continuities are a kind of smoothness conditions for functions. 

\subsubsection*{Continuity of Existing Generalizations.}

The continuity class that a generalized clustering coefficient belongs to depends on what kind of functions the coefficient is composed of.

Because a H\"{o}lder continuous function $ (\alpha < 1)$ is more weakly continuous than a Lipschitz continuous function, a generalized clustering coefficient consisted of H\"{o}lder continuous functions is a little more sensitive with small changes of edge weights than one consisted of Lipschitz continuous functions. In addition, between generalized clustering coefficients consisted of Lipschitz continuous functions, one that has larger Lipschitz constant\footnote{A function $f$ is called Lipschitz continuous in $D$ if there exists a real constant $K$ such that $|f(x) - f(y)| < K |x - y|$ for all $x$ and $y$ in $D$. Any such $K$ is referred to as a Lipschitz constant for the function $f$.} is more sensitive than the other that has smaller one. 

For example, the two-arguments functions 'multiplication' and 'maximum' are Lipschitz continuous. A composed function of Lipschitz continuous functions has also strong continuity. In particular, the generalizations by Zhang et al. (Eq. (\ref{zhang}) and (\ref{kalna})) and Holme et al. (Eq. (\ref{holme})) consisted of Lipschitz continuous functions.

To minimize the error of the coefficient value, it may be preferable to use a generalized clustering coefficient that has stronger continuity.

\section{Continuous Generalizations of Clustering Coefficient}
\label{section3}

In this section, we propose our generalized clustering coefficients for weighted and directed networks. First we propose generalizations of the local clustering coefficient. Second we propose generalizations of the global clustering coefficient. We summarize their properties in the end of this section.

\subsection{Our Generalizations of Local Clustering Coefficient}
\begin{figure}[htbp]
 \begin{center}
  \includegraphics[width=7.5cm,bb=0 0 1761 411]{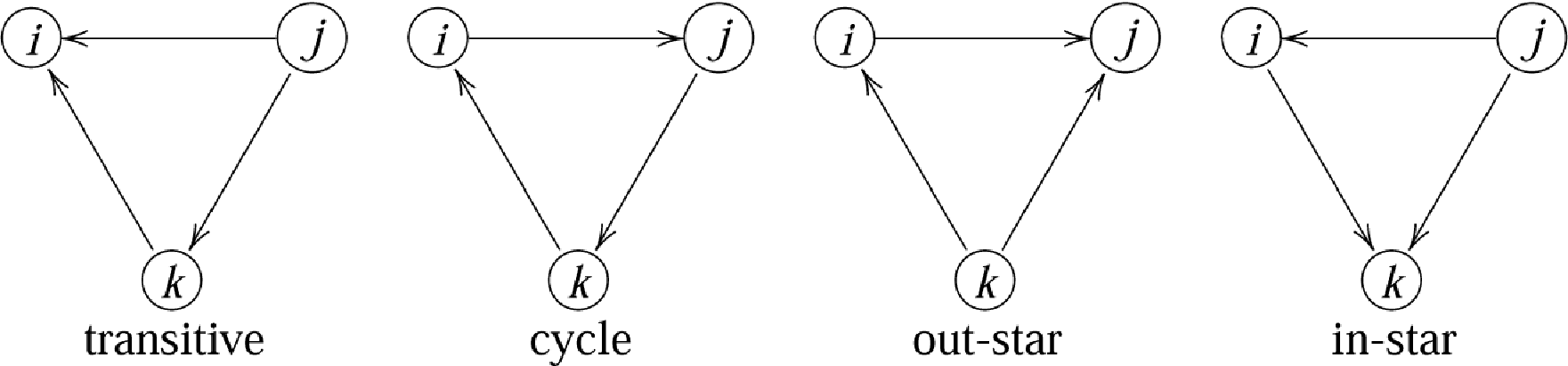}
 \end{center}
 \caption{Four configurations of a triplet.}
 \label{fig:configurations}
\end{figure}

First, we consider how to generalize the local clustering coefficient to weighted and directed networks. The definition of the local clustering coefficient (Eq.(\ref{watts})) can be rewritten with the weights and directionality of edges as follows\footnote{The similar rewriting was used in the generalizations for weighted and undirected networks \cite{grindrod2002range,zhang2005general,kalna2006clustering,holme2007korean}.}:
\begin{eqnarray}\label{localteian1}
C(k) = \frac{\displaystyle\sum_{i,j} \; w_{k \rightarrow i} \cdot w_{k \rightarrow j} \cdot w_{i \rightarrow j}}{\displaystyle\sum_{i,j} \; w_{k \rightarrow i} \cdot w_{k \rightarrow j} \cdot \displaystyle \mathrm{max}_{i,j}(w_{i \rightarrow j})}.
\end{eqnarray}
In directed networks, the possible patterns of a triplet consist of four configurations (Figure \ref{fig:configurations}). 
In Eq.(\ref{localteian1}), we choose the 'out-star' configuration as a concrete example.
Eq.(\ref{localteian1}) can also be rewritten with a certain two-argument function $h_m(x,y)=x \cdot y$. 
We define the generalized clustering coefficient as follows:
\begin{eqnarray}\label{localteian2}
C(k)_{gloc} = \frac{\displaystyle {\scriptstyle \sum_{i,j} \; h_m(\, h_m(w_{k \rightarrow i},\; w_{k \rightarrow j}), \; w_{i \rightarrow j})}}{\displaystyle {\scriptstyle \sum_{i,j} \; h_m(\, h_m(w_{k \rightarrow i}, \;w_{k \rightarrow j}),\;\mathrm{max}_{i,j}(w_{i \rightarrow j}))}}.
\end{eqnarray}
Here we can replace $h_m$ by a binary function $h$ that satisfies the following requirements:
\begin{eqnarray}
h(x,y) = 
\begin{cases}
0 &\textrm{if } x = y = 0 ,\\
0 &\textrm{if } x = 0 \textrm{ and } y = 1 ,\\
0 &\textrm{if } x = 1 \textrm{ and } y = 0 ,\\
1 &\textrm{if } x = y = 1 .\\
\end{cases} 
\end{eqnarray}
Here we use a continuous function $h$. Because composition of continuous functions is again continuous, the generalized clustering coefficient comes to be continuous.

More generally, we consider a ternary function $g$ instead of $h$ as follows:
\begin{eqnarray}\label{localteian3}
C(k)_{gloc} &=& \frac{\scriptstyle\sum_{i,j} \; g (\, w_{k \rightarrow i},\; w_{k \rightarrow j}, \; w_{i \rightarrow j})}{\scriptstyle \sum_{i,j} \; g(\, w_{k \rightarrow i}, \;w_{k \rightarrow j}, \;\mathrm{max}_{i,j}(w_{i \rightarrow j}))}.\\
g(x,y,z) &=& 
\begin{cases}
0 &\textrm{if one of} \; x , \; y , \; z \; \textrm{is} \; 0 ,\\
1 &\textrm{if } x = y = z = 1 .\\
\end{cases} \label{eqg} 
\end{eqnarray}
However, we consider binary cases in the sequel. As examples of binary functions, we investigate four methods (multiplication, geometric mean, minimum, harmonic mean). 

\subsubsection*{Multiplication}
We define the generalization of the local clustering coefficient with $h(x,y) = xy$ as follows:
\begin{eqnarray}
C_{loc,mul}(k) = \frac{\displaystyle\sum_{i,j} \; w_{k \rightarrow i} \cdot w_{k \rightarrow j} \cdot w_{i \rightarrow j}}{\displaystyle\sum_{i,j} \; w_{k \rightarrow i} \cdot w_{k \rightarrow j} \cdot \mathrm{max}_{i,j}(w_{i \rightarrow j}) }.
\end{eqnarray}
The similar definition was already proposed as weighted and undirected networks \cite{zhang2005general,holme2007korean}. 

\subsubsection*{Geometric Mean}
We define the generalization of the local clustering coefficient by letting $h(x,y) = \sqrt{xy}$ as follows:
\begin{eqnarray}
C_{loc,gm}(k) = \frac{\displaystyle\sum_{i,j} \; \sqrt{ \; \sqrt{ w_{k \rightarrow i} \cdot w_{k \rightarrow j}} \cdot w_{i \rightarrow j} }}{\displaystyle\sum_{i,j} \; \sqrt{ \sqrt{  w_{k \rightarrow i} \cdot w_{k \rightarrow j} } \cdot \mathrm{max}_{i,j}(w_{i \rightarrow j}) }}.
\end{eqnarray}

\subsubsection*{Minimum}
We define the generalization of the local clustering coefficient by letting $h(x,y) = \mathrm{min}(x,y)$ as follows:
\begin{eqnarray}
C_{loc,min}(k) = \frac{\displaystyle {\scriptstyle \sum_{i,j} \; \mathrm{min}( \; \mathrm{min}( w_{k \rightarrow i}, \; w_{k \rightarrow j}),\; w_{i \rightarrow j})}}{\displaystyle {\scriptstyle \sum_{i,j} \; \mathrm{min}( \mathrm{min} (w_{k \rightarrow i}, \; w_{k \rightarrow j}) ,\; \mathrm{max}_{i,j}(w_{i \rightarrow j}) )}}.
\end{eqnarray}

\subsubsection*{Harmonic Mean}
We define the generalization of the local clustering coefficient with $h(x,y) = \frac{2}{\frac{1}{x} + \frac{1}{y}}$ as follows:
\begin{eqnarray}
C_{loc,har}(k) = \displaystyle \frac{\displaystyle\sum_{i,j} \; \left( \frac{2}{  \frac{1}{ \frac{2}{ \frac{1}{w_{k \rightarrow i}} + \frac{1}{w_{k \rightarrow j}} }} + \frac{1}{w_{i \rightarrow j}}} \right)} {\displaystyle\sum_{i,j} \; \left( \frac{2}{ \frac{1}{ \frac{2}{ \frac{1}{w_{k \rightarrow i}} + \frac{1}{w_{k \rightarrow j}} }} + \frac{1}{\mathrm{max}_{i,j}(w_{i \rightarrow j})}} \right)}.
\end{eqnarray}

\subsubsection{Brief Comparison of Generalizations}\label{compare1}
Here, we briefly compare the existing and our generalizations of the local clustering coefficient for weighted and directed networks.

\begin{figure}[htbp]
 \begin{center}
  \includegraphics[width=8cm,bb=0 0 384 230]{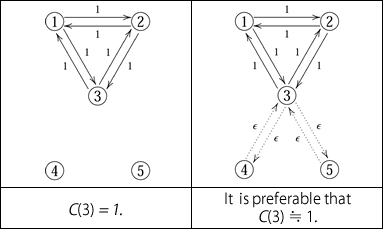}
 \end{center}
 \caption{Two similar simple networks.}
 \label{fig:example5nodes}
\end{figure}

Suppose there are two networks as Figure \ref{fig:example5nodes} shows. Both networks consist of 5 nodes. In the left-hand network, there are the bilateral edges of weight 1 among node 1 , node 2 and node 3. Node 4 and Node 5 are isolated nodes. In the right-hand network, in addition to the bilateral edges of weight 1 among 3 nodes, node 4 and node 5 have the bilateral edges of weight $\epsilon$ to node 3 ($\epsilon$ is near zero value).
If $\epsilon$ is vanishingly low, one may consider that the two networks are almost the same in practice.
In that case, it is preferable that the coefficient values of two networks are close. In particular, $C(1)$ (the coefficient value of node 1) and $C(2)$ and $C(3)$ equal 1 in the left-hand network. In the right-hand network, although $C(1)$ and $C(2)$ are also 1, there is a difference in the value of $C(3)$.

Table \ref{briefcomparelocal} shows the almost values of $C(3)$ in the
right-hand network with each of the generalized clustering coefficient
when $\epsilon = 0.0001$.

In the case of the discontinuous generalized clustering coefficients, the values of $C(3)$ in the right-hand network are far from 1. On the other hand, the values of $C(3)$ by our continuous clustering coefficients are almost 1.

\begin{table}[htbp]
\begin{center}
\caption{Almost values of $C(3)$ in the right-hand network.}
\label{briefcomparelocal}
\begingroup
\renewcommand{\arraystretch}{1.3}
\footnotesize
\begin{tabular}{|l|c|} \hline
(Existing) & $C(3)$\\\hline
$C(3)_{fag}$		& 0.1667 (16.68\%) \\
$C(3)_{suz}$		& 0.0357 (3.57\%)\\\hline
(Proposed) & $C(3)$\\\hline
$C(3)_{loc,mul}$ 	& 0.9996 (100.0\%) \\
$C(3)_{loc,gm}$ 	& 0.7092 (70.9\%)\\
$C(3)_{loc,min}$ 	& 0.9995 (100.0\%) \\
$C(3)_{loc,har}$ 	& 0.9982 (99.8\%)\\\hline
\end{tabular}
\endgroup
\end{center}
\end{table}

Here we define each generalization with the 'out-star' configuration. Without going into further detail, we mention that there are generalizations with other configurations. In some cases, it may be advantageous to use other combination of the definitions with more than one or all configurations.

\subsection{Our Generalizations of Global Clustering Coefficient}
We now consider how to generalize the global clustering coefficient to weighted and directed networks. Our method was inspired by the method by Opsahl et al. \cite{opsahl2009clustering} (see section \ref{opsahl1} and \ref{opsahl2}). We also adopt their method that calculates the values of triplets. Our method differs from theirs in distinguishing between the values of closed triplets and the values of triplets (open triplets and closed triplets). In addition, we do not distinguish the configurations of triplets\footnote{In the method by Opsahl et al., the values of triplets are calculated without considering vacuous triplets, and non-vacuous and intransitive closed triplets \cite{opsahl2009clustering}.}. 
We define generalized clustering coefficients as follows:
\begin{eqnarray}\label{globalteian1}
C_{gglo} &=& \frac{\rm{total\:of\:closed\:triplets}}{\rm{total\:of\:triplets}}  \notag \\
&=& \frac{\displaystyle \sum_{\tau_\Delta} \omega_\Delta}{\displaystyle \sum_{\tau} \omega}. 
\end{eqnarray}
Here $\tau$ is the set of triplets and $\tau_\Delta$ is the set of closed triplets. $\omega$ is the value of a triplet and $\omega_\Delta$ is the value of a closed triplet.  

The total of triplets is calculated separately for each configuration with a ternary function $g$ in Eq.(\ref{eqg}) as follows:
\begin{align}\label{omega}
\sum_{\tau} \omega &=
\displaystyle \sum_{i,j} \; g(w_{j \rightarrow k}, \;w_{k \rightarrow i}, \;\mathrm{max}_{i,j}(w_{i \rightarrow j})) \notag \\
 \quad & + \displaystyle\sum_{i,j} \; g(w_{j \rightarrow k}, \;w_{k \rightarrow i}, \;\mathrm{max}_{i,j}(w_{i \rightarrow j})) \notag \\
 \quad & + \displaystyle\sum_{i,j} \; g(w_{k \rightarrow i}, \;w_{k \rightarrow j}, \;\mathrm{max}_{i,j}(w_{i \rightarrow j})) \notag \\
 \quad & + \displaystyle\sum_{i,j} \; g(w_{i \rightarrow k}, \;w_{j \rightarrow k}, \;\mathrm{max}_{i,j}(w_{i \rightarrow j})).
\end{align}
The total value of closed triplets is calculated separately for each configuration with $g$:
\begin{align}\label{omega_delta}
\sum_{\tau_\Delta} \omega_\Delta &= 
\displaystyle\sum_{i,j} \; g(w_{j \rightarrow k}, \;w_{k \rightarrow i}, \;w_{j \rightarrow i}) \notag \\
\quad & + \displaystyle\sum_{i,j} \; g(w_{j \rightarrow k}, \;w_{k \rightarrow i}, \;w_{i \rightarrow j}) \notag \\
\quad & + \displaystyle\sum_{i,j} \; g(w_{k \rightarrow i}, \;w_{k \rightarrow j}, \;w_{i \rightarrow j}) \notag \\
\quad & + \displaystyle\sum_{i,j} \; g(w_{i \rightarrow k}, \;w_{j \rightarrow k}, \;w_{j \rightarrow i}).
\end{align}
Here we use a continuous function $g$. Thus the generalized
coefficient comes to be continuous.
\begin{table}[htbp]
\begin{center}
\caption{Triplet value $\omega$ and closed triplet value $\omega_\Delta$.}
\label{tripletsvalueexample}
\begingroup
\renewcommand{\arraystretch}{1.3}
\footnotesize
\begin{tabular}{|l|c|c|c|c|} \hline
& \multicolumn{2}{|c|}{\includegraphics[width=1.7cm,bb=0 0 150 150]{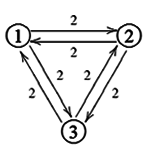}} & \multicolumn{2}{|c|}{\includegraphics[width=1.7cm,bb=0 0 150 150]{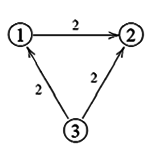}} \\ \hline\hline
(Existing)  &\makebox[25pt][c]{ $\omega$} &\makebox[25pt][c]{ $\omega_\Delta$} &\makebox[25pt][c]{ $\omega$} & \makebox[25pt][c]{$\omega_\Delta$} \\ \hline 
$C_{ops,am}$	& 12 & 12 & 2 & 2 \\ \hline
$C_{ops,gm}$	&  12 & 12 & 2 & 2 \\ \hline
$C_{ops,min}$	&  12 & 12 & 2 & 2 \\ \hline
$C_{ops,max}$	& 12 & 12 & 2 & 2 \\ \hline
(Proposed)  &\makebox[25pt][c]{ $\omega$} &\makebox[25pt][c]{ $\omega_\Delta$} &\makebox[25pt][c]{ $\omega$} & \makebox[25pt][c]{$\omega_\Delta$} \\ \hline 
$C_{glo,mu}$	&  192 & 192 & 48 & 24 \\ \hline
$C_{glo,gm}$	&  48 & 48 & 12 & 6 \\ \hline
$C_{glo,min}$	&  48 & 48 & 12 & 6 \\ \hline
$C_{glo,hm}$	& 48 & 48 & 12 & 6 \\ \hline
\end{tabular}
\endgroup
\end{center}
\end{table}

As examples of ternary functions,
we propose four methods (multiplication, geometric mean, minimum,
harmonic mean) for calculating $\omega$ and $\omega_\Delta$.

Table \ref{tripletsvalueexample} shows examples of calculating $\omega$ and $\omega_\Delta$.
In the left triplet, all values of $C_{ops}$ and $C_{gglo}$ are 1. In the right triplet, all values of $C_{ops} = 1$ and all values of $C_{gglo} = 0.5$.

\subsubsection{Brief Comparison of Generalizations}
Here we briefly compare the existing and our generalizations of the global clustering coefficient for weighted and directed networks.

In the same way of Section \ref{compare1}, we calculate the coefficient values of the two networks in Figure \ref{fig:example5nodes}. Table \ref{briefcompareglobal} shows the coefficient values of the right-hand network when $\epsilon = 0.0001$.
The values of $C$ of the right-hand network are far from 1 when the methods 'arithmetic mean' and 'maximum' of $C_{ops}$ are adapted. On the other hand, the values of our generalized clustering coefficients are nearly 1. Although the values of $C$ by the methods 'geometric mean' and 'minimum' of $C_{ops}$ are nearly 1 and they seem to be no problem, we show a problematic case in the next section.

Here we define the generalizations of the global clustering coefficient with the summation of the values of the triplets of each configuration. Without going into further detail, we mention that there are generalizations with other configurations. In some cases, it may be advantageous to use other combinations of configurations.

\begin{table}[htbp]
\begin{center}
\caption{Values of the clustering coefficient of the right-hand network.}
\label{briefcompareglobal}
\begingroup
\renewcommand{\arraystretch}{1.3}
\footnotesize
\begin{tabular}{|l|c|} \hline
(Existing) & $C$ \\ \hline
$C_{ops,am}$ & 0.6000 (60.0\%) \\
$C_{ops,gm}$ & 0.9868 (98.7\%) \\
$C_{ops,min}$ & 0.9998 (100.0\%) \\
$C_{ops,max}$ & 0.4286 (42.9\%)\\\hline
(Proposed) & $C$ \\ \hline
$C_{glo,mu}$ & 0.9999 (100.0\%) \\
$C_{glo,gm}$ & 0.9411 (94.1\%)\\
$C_{glo,min}$ & 0.9998 (100.0\%)\\
$C_{glo,hm}$ & 0.9996 (100.0\%)\\\hline
\end{tabular}
\endgroup
\end{center}
\end{table}

\subsection{Properties of Generalized Clustering Coefficients}

Our generalizations are designed to admit the following requirements:
Continuity, General versatility, weight-scale invariance (Table \ref{teiantable}).

\newcommand{\ms}{\hspace*{-.3em}}

\begin{table}[htbp]
\begin{center}
\caption{Properties of our clustering coefficients.}
\label{teiantable}
\begingroup
\renewcommand{\arraystretch}{1.3}
\footnotesize
\begin{tabular}{|l|c|c|c|} \hline
             & \ms Continuity \ms & \ms General Versatility \ms & \ms \shortstack{Weight-Scale \\ Invariance} \ms \rule[0mm]{0mm}{6mm} \\\hline
$C(k)_{gloc}$ & X & X & X \\\hline
$C_{gglo}$ & X & X & X \\\hline
\end{tabular}
\endgroup
\end{center}
\end{table}

The measure of the continuity of each of the generalized clustering coefficient is different depending on the choice of the function $h$ and $g$. 'Multiplication' , 'Minimum' and 'Harmonic Mean' are Lipschitz continuous. 'Geometric Mean' is H\"{o}lder continuous. Therefore, $C_{loc,gm}(k)$ and $C_{glo,gm}$ is more sensitive than the other generalizations of the local or global clustering coefficients.

\section{Experiments}
\label{section4}

In this section, we describe the method of numerical experiments.
Here we limited the objects of the experiments to examining the generalizations of the local and global clustering coefficients for weighted and directed networks.
We examined the robustness against edge weight errors of our continuous generalized clustering coefficients by experiments on artificial network models and real world network datasets.

All experiments were performed using R\footnote{http://www.r-project.org/} version 2.9.2 for Windows.

\subsection{Artificial Network Datasets}

In this subsection, we describe the experiments on two artificial network models.
Model 1 is a network for examining the existing and our generalizations of the local clustering coefficient. Model 2 is a network for examining the generalizations of the global clustering coefficient.

\subsubsection{Model 1}\label{subsecmodel1}
\begin{figure}[htbp]
 \begin{center}
  \includegraphics[width=7cm,bb=0 0 360 420]{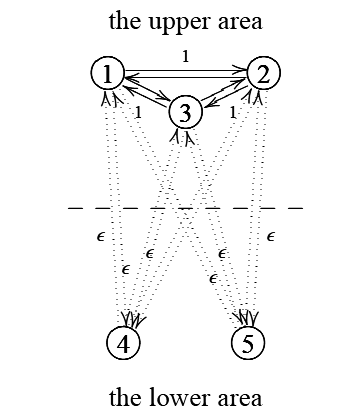}
 \end{center}
 \caption{Artificial Network: Model 1.}
 \label{fig:model1}
\end{figure}

A network Model 1 consists of two parts (Figure \ref{fig:model1}): the upper area and the lower area. There are bilateral edges of weight 1 among the all nodes in the upper area. And there are bilateral edges of weight $\epsilon$ between the all nodes in the upper area and the all nodes in the lower area.

In model 1, we examined changes of coefficient values when $\epsilon$ is reduced from 1 to 0. In case that $\epsilon = 0$, although the local generalized clustering coefficients of the nodes in the lower area are undefined because they are isolated nodes, the coefficients in the upper area are defined. Note that the coefficients of the upper area are also undefined when the number of the nodes in the upper area is less than 2. We calculated the local coefficient values of the nodes of the upper area. 

We let the pairs of the numbers of the nodes in the upper area and lower area be (40, 4), (8, 8) and (4, 40). 

\subsubsection{Model 2}\label{subsecmodel2}
\begin{figure}[htbp]
 \begin{center}
  \includegraphics[width=7cm,bb=0 0 360 420]{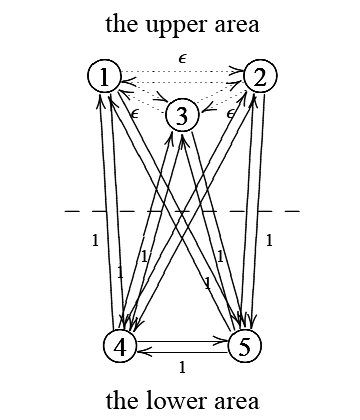}
 \end{center}
 \caption{Artificial Network: Model 2.}
 \label{fig:model2}
\end{figure}

A network Model 2 consists of two parts (Figure \ref{fig:model2}) as in Section \ref{subsecmodel1}. In this network, there are bilateral edges of weight $\epsilon$ among the all nodes in the upper area. And there are bilateral edges of weight 1 among the all nodes in the lower area. And there are bilateral edges of weight 1 between the all nodes of the upper area and the all nodes in the lower area.

In model 2, we examined changes of coefficient values when $\epsilon$ is reduced from 1 to 0. In case that $\epsilon = 0$, if there are two nodes in the upper area and one node in the lower area, the global generalized clustering coefficients are defined. We calculated the coefficient values in the whole network. 

We let the pairs of the numbers of the nodes in the upper area and lower area be (40, 4), (8, 8) and (4, 40) respectively.


\subsection{Real World Network Datasets}

In this subsection, we describe the experiments on real world network datasets. 

To examine the robustness of our continuous clustering coefficients, we focused on the situation that there are some measurement errors of edge weights in a network. In such a situation, the measured edge weights differ from the actual values. Thus in the case that one calculates a discontinuous clustering coefficient, the coefficient value may significantly differ from the actual value.

\begin{figure}[htbp]
 \begin{center}
  \includegraphics[width=8cm,bb=0 0 634 466]{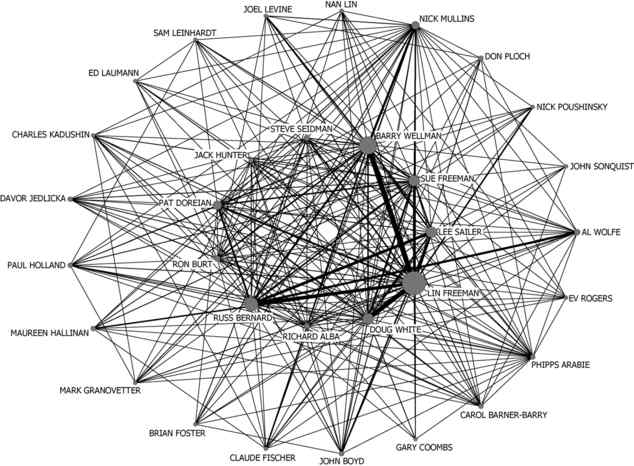}
 \end{center}
 \caption{Freeman's EIES network. The size of a node is proportional to the number of the messages sent by the researcher, and the width of an edge between two nodes corresponds to the number of the messages exchanged between the two researchers. (cited from Opsahl et al. \cite{opsahl2009clustering})}
 \label{fig:freeman}
\end{figure}

We conducted the experiments with two datasets. The first dataset is Freeman's EIES network \cite{freeman1979networkers}.
This network is also used by Wasserman et al. \cite{wasserman1994social} and Opsahl et al. \cite{opsahl2009clustering}.
This network is represented by a frequency matrix of the number of the messages sent among 32 researchers that electronically communicate.
The average of the edge weights is 33.7. The minimum weight is 2 except for weight 0. The maximum weight is 559.
Figure \ref{fig:freeman} shows Freeman's EIES network.

The second dataset is a neural network of Caenorhabditis elegans worm (C.elegans). This network is used by Watts et al. \cite{watts1998collective}, and used by Opsahl et al. \cite{opsahl2009clustering}. This network consists of 306 neurons. Two neurons are connected if at least one synapse or gap junction exist between them. The weight of an edge represents the number of synapses and gap junctions.
The average of the edge weights is 3.76. The minimum weight is 1 except for weight 0. The maximum weight is 70.

For the above datasets, we added errors to the weights for all edges and compared our generalizations of the global clustering coefficient with the generalizations by Opsahl et al. \cite{opsahl2009clustering}. 

We conducted two experiments by two methods of error addition.

\subsubsection{Experiment 1}

In this experiments, we added a minute amount of uniform positive values to the edge weights. We added the weight 0 (not changed) , $10^{-12}$, $10^{-9}$ and $10^{-6}$, and calculated the coefficient values respectively. As the result of additions, each network became a complete graph.

\subsubsection{Experiment 2}

In this experiments, we added normal random numbers with mean and standard deviation = 0 (not changed) , $10^{-12}$, $10^{-9}$, $10^{-6}$ to the edge weights respectively. In the case of negative weights, we treat them as weight 0. We calculated the mean of 10 trials.

\section{Results}
\label{section5}

In this section, we describe the results of the numerical experiments.

\subsection{Artificial Network Datasets}

In this subsection, we describe the results of the experiments on the two artificial networks.

\subsubsection{Model 1}\label{subsecresultmodel1}

\begin{figure}[htbp]
 \begin{center}
  \includegraphics[width=8cm,bb=0 0 2688 2300]{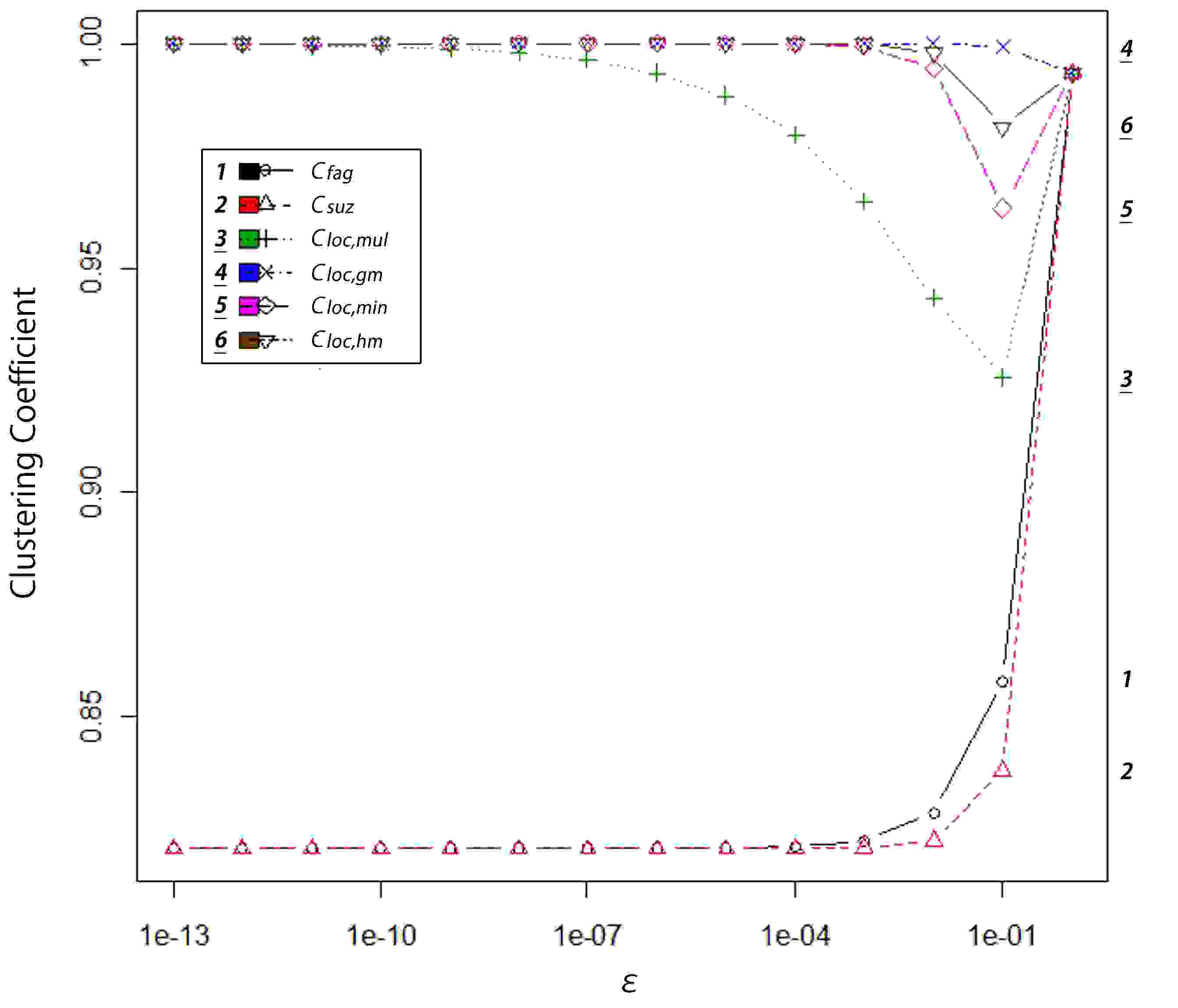}
 \end{center}
 \caption{Values of the generalized clustering coefficients of the nodes in the upper area (Model 1: upper = 40 nodes, lower = 4 nodes).}
 \label{fig:model1result40-4}
\end{figure}

\begin{figure}[htbp]
 \begin{center}
  \includegraphics[width=8cm,bb=0 0 2688 2300]{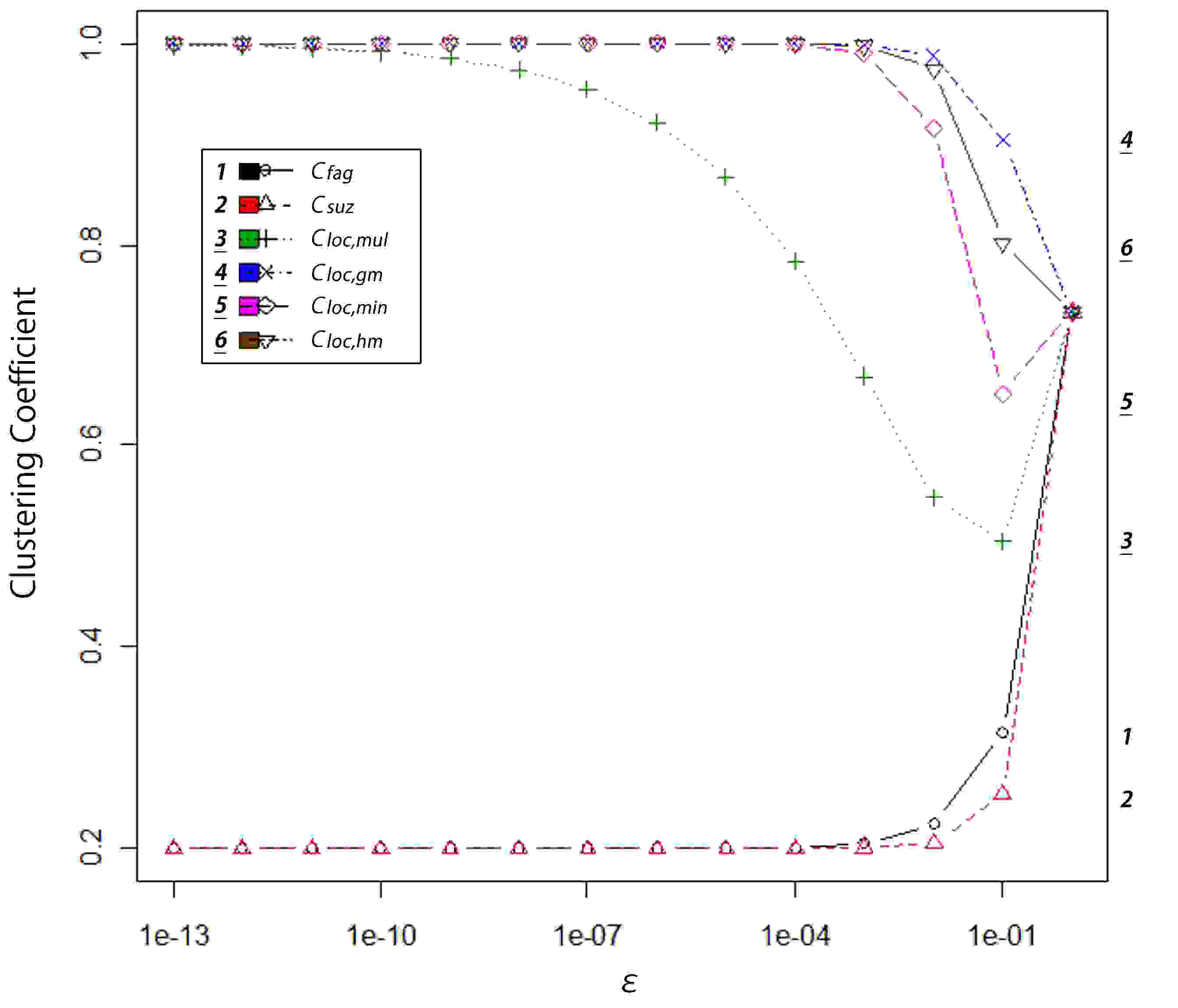}
 \end{center}
 \caption{Values of the generalized clustering coefficients of the nodes in the upper area (Model 1: upper = 8 nodes, lower = 8 nodes).}
 \label{fig:model1result8-8}
\end{figure}

\begin{figure}[htbp]
 \begin{center}
  \includegraphics[width=8cm,bb=0 0 2688 2300]{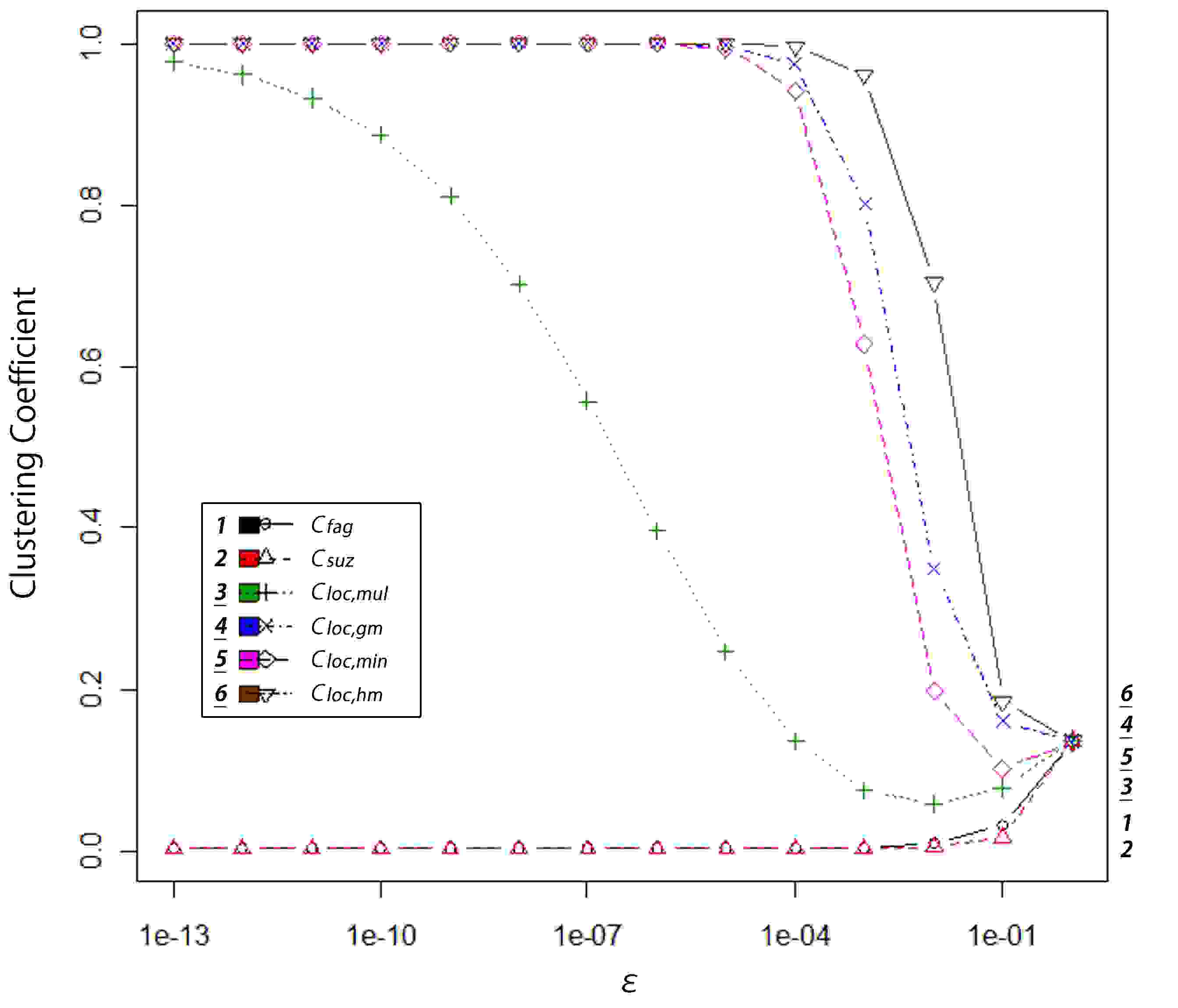}
 \end{center}
 \caption{Values of the generalized clustering coefficients of the nodes in the upper area (Model 1: upper = 4 nodes, lower = 40 nodes).}
 \label{fig:model1result4-40}
\end{figure}

The values of the generalized clustering coefficients of the nodes in the upper area are plotted in Figure \ref{fig:model1result40-4}, Figure \ref{fig:model1result8-8} and Figure \ref{fig:model1result4-40} respectively. Table \ref{model1table40-4}, Table \ref{model1table8-8} and Table \ref{model1table4-40} show the values of the generalized clustering coefficients of the upper area in the case that $\epsilon$ is nearly zero.

\begin{table*}[htbp]
\begin{center}
\caption{Values of the generalized clustering coefficients of the nodes in the upper area near weight 0 (Model 1: upper = 40 nodes, lower = 4 nodes).}
\label{model1table40-4}
\begingroup
\renewcommand{\arraystretch}{1.3}
\scriptsize
\begin{tabular}{|l|c|c|c|c|} \hline
(Existing) & $\epsilon = 0$ & $\epsilon = 1^{-13}$ & $\epsilon = 1^{-12}$ & $\epsilon = 1^{-11}$ \rule[0mm]{0mm}{2.7mm} \\\hline
$C_{fag}(k)$ & 1 (100.0\%) & 0.8206 (82.1\%) & 0.8206 (82.1\%) & 0.8206 (82.1\%)  \\
$C_{suz}(k)$	& 1 (100.0\%) & 0.8206 (82.1\%) & 0.8206 (82.1\%) & 0.8206 (82.1\%)  \\\hline
(Proposed) & $\epsilon = 0$ & $\epsilon = 1^{-13}$ & $\epsilon = 1^{-12}$ & $\epsilon = 1^{-11}$ \rule[0mm]{0mm}{2.7mm} \\\hline
$C_{loc,mul}(k)$ & 1 (100.0\%) & 1 (100.0\%) & 1 (100.0\%) & 1 (100.0\%) \\
$C_{loc,gm}(k)$ & 1 (100.0\%) & 0.9999 (100.0\%) & 0.9998 (100.0\%) & 0.9996 (100.0\%) \\
$C_{loc,min}(k)$ & 1 (100.0\%) & 1 (100.0\%) & 1 (100.0\%) & 1 (100.0\%) \\
$C_{loc,har}(k)$ & 1 (100.0\%) & 1 (100.0\%) & 1 (100.0\%) & 1 (100.0\%) \\\hline
\end{tabular}
\endgroup
\end{center}
\end{table*}

\begin{table*}[htbp]
\begin{center}
\caption{Values of the generalized clustering coefficients of the nodes in the upper area near weight 0 (Model 1: upper = 8 nodes, lower = 8 nodes).}
\label{model1table8-8}
\begingroup
\renewcommand{\arraystretch}{1.3}
\scriptsize
\begin{tabular}{|l|c|c|c|c|} \hline
(Existing) & $\epsilon = 0$ & $\epsilon = 1^{-13}$ & $\epsilon = 1^{-12}$ & $\epsilon = 1^{-11}$ \\\hline
$C_{fag}(k)$	& 1 (100.0\%) & 0.2 (20\%) & 0.2 (20\%) & 0.2 (20\%) \\
$C_{suz}(k)$	& 1 (100.0\%) & 0.2 (20\%) & 0.2 (20\%) & 0.2 (20\%) \\\hline
(Proposed) & $\epsilon = 0$ & $\epsilon = 1^{-13}$ & $\epsilon = 1^{-12}$ & $\epsilon = 1^{-11}$ \\\hline
$C_{loc,mul}(k)$	& 1 (100.0\%) & 1 (100.0\%) & 1 (100.0\%) & 1 (100.0\%) \\
$C_{loc,gm}(k)$	& 1 (100.0\%) & 0.9985 (99.9\%) & 0.9973 (99.7\%) & 0.9953 (99.5\%) \\
$C_{loc,min}(k)$	& 1 (100.0\%) & 1 (100.0\%) & 1 (100.0\%) & 1 (100.0\%) \\
$C_{loc,har}(k)$	& 1 (100.0\%) & 1 (100.0\%) & 1 (100.0\%) & 1 (100.0\%) \\\hline
\end{tabular}
\endgroup
\end{center}
\end{table*}

\begin{table*}[htbp]
\begin{center}
\caption{Values of the generalized clustering coefficients of the nodes in the upper area near weight 0 (Model 1: upper = 40 nodes, lower = 4 nodes).}
\label{model1table4-40}
\begingroup
\renewcommand{\arraystretch}{1.3}
\scriptsize
\begin{tabular}{|l|c|c|c|c|} \hline
(Existing) & $\epsilon = 0$ & $\epsilon = 1^{-13}$ & $\epsilon = 1^{-12}$ & $\epsilon = 1^{-11}$ \\\hline
$C_{fag}(k)$	& 1 (100.0\%) & 0.0033 (0.33\%) & 0.0033 (0.33\%)  & 0.0033 (0.33\%) \\
$C_{suz}(k)$	& 1 (100.0\%) &  0.0033 (0.33\%)  &  0.0033 (0.33\%)  &  0.0033 (0.33\%) \\\hline
(Proposed) & $\epsilon = 0$ & $\epsilon = 1^{-13}$ & $\epsilon = 1^{-12}$ & $\epsilon = 1^{-11}$ \\\hline
$C_{loc,mul}(k)$ 	& 1 (100.0\%) & 1 (100.0\%) & 1 (100.0\%) & 1 (100.0\%) \\
$C_{loc,gm}(k)$ 	& 1 (100.0\%) &  0.9779 (97.8\%) &   0.9613 (96.1\%) & 0.9329 (93.3\%) \\
$C_{loc,min}(k)$ 	& 1 (100.0\%) & 1 (100.0\%) & 1 (100.0\%) & 1 (100.0\%) \\
$C_{loc,har}(k)$ 	& 1 (100.0\%) & 1 (100.0\%) & 1 (100.0\%) & 1 (100.0\%) \\\hline
\end{tabular}
\endgroup
\end{center}
\end{table*}

In model 1, it is preferable that the coefficient values of the nodes of the upper area close to 1 when $\epsilon$ is reduced from 1 to 0. However, in the case of the generalization by Fagiolo \cite{fagiolo2007clustering} and Suzuki \cite{suzuki2009fukuzatsu}, the coefficient values did not approach to 0. On the other hand, the values by the proposed coefficients approached to 1.

\subsubsection{Model 2}\label{subsecresultmodel2}

\begin{figure}[htbp]
 \begin{center}
  \includegraphics[width=8cm,bb=0 0 2688 2300]{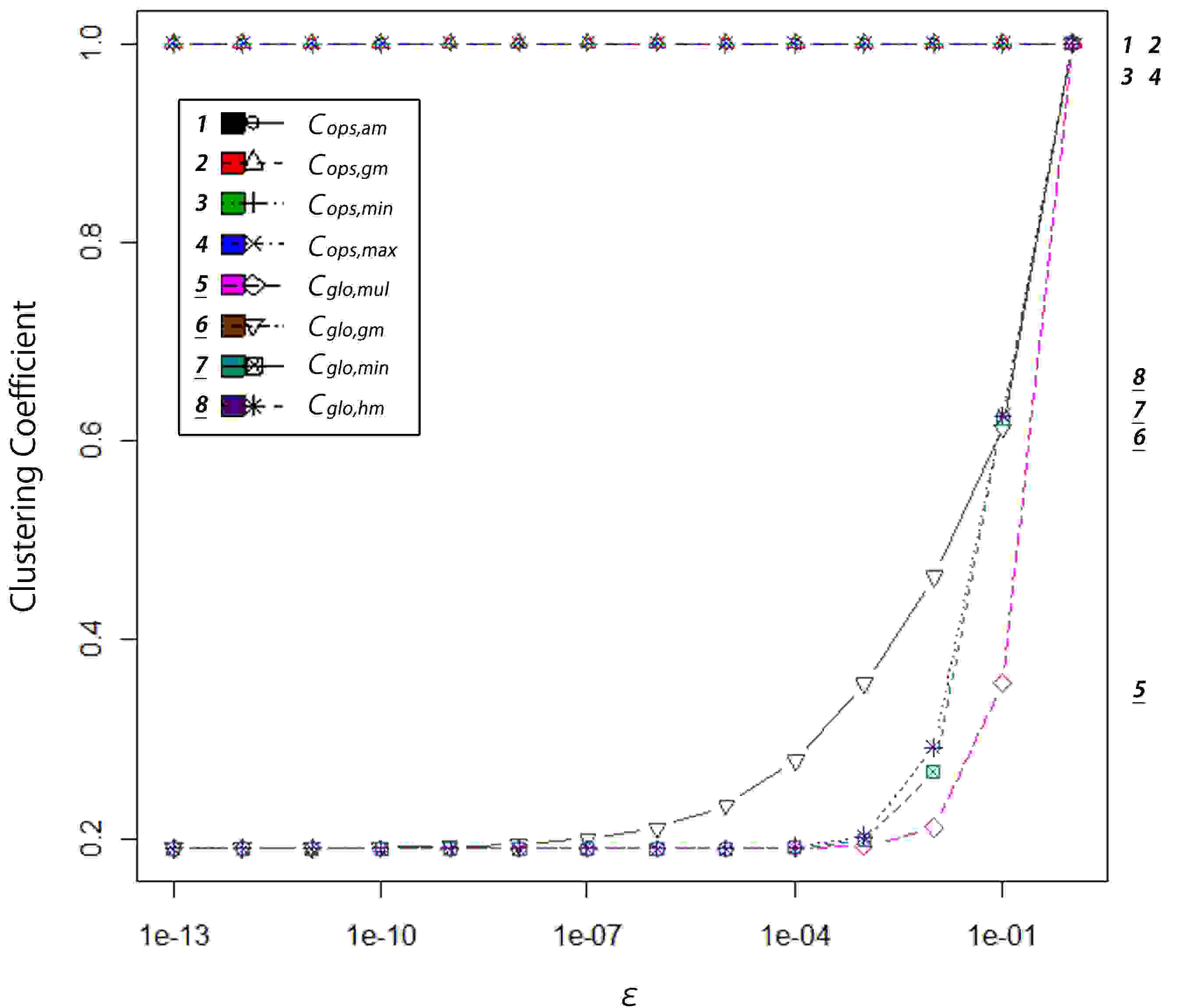}
 \end{center}
 \caption{Values of the generalized clustering coefficients of the whole network (Model 2: upper = 40 nodes, lower = 4 nodes).}
 \label{fig:model2result40-4}
\end{figure}

\begin{figure}[htbp]
 \begin{center}
  \includegraphics[width=8cm,bb=0 0 2688 2300]{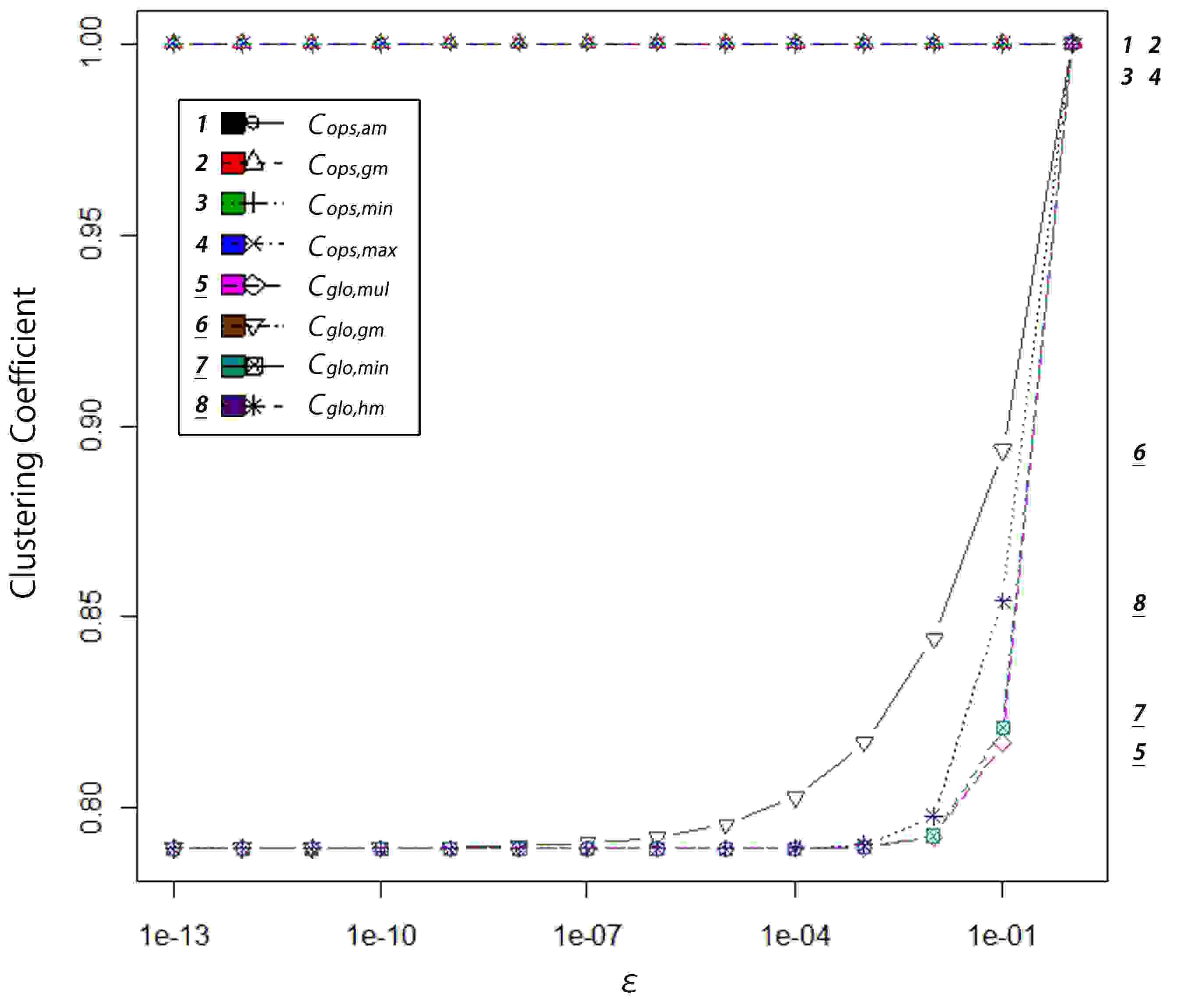}
 \end{center}
 \caption{Values of the generalized clustering coefficients of the whole network (Model 2: upper = 8 nodes, lower = 8 nodes).}
 \label{fig:model2result8-8}
\end{figure}

\begin{figure}[htbp]
 \begin{center}
  \includegraphics[width=8cm,bb=0 0 2688 2300]{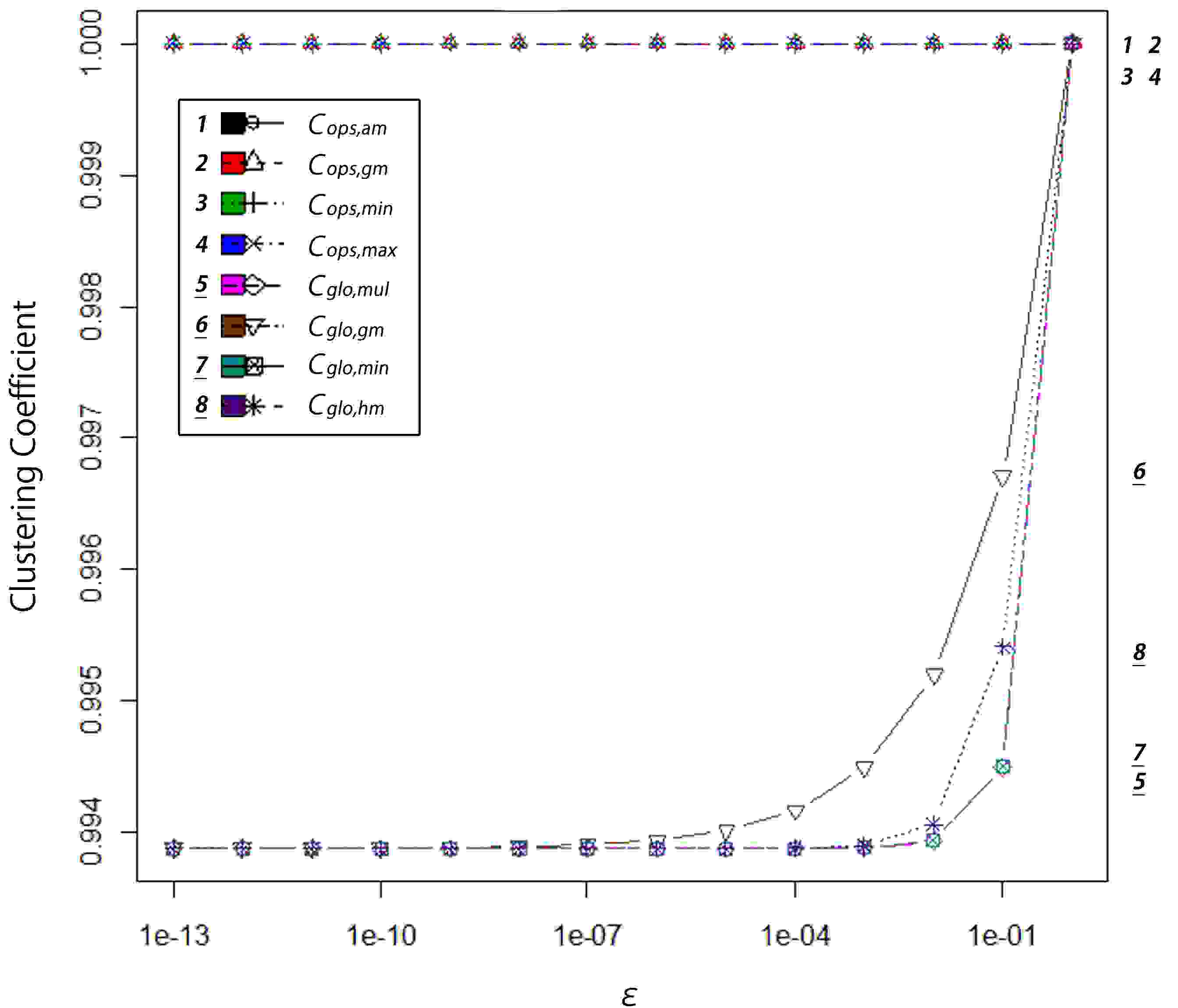}
 \end{center}
 \caption{Values of the generalized clustering coefficients of the whole network (Model 2: upper = 4 nodes, lower = 40 nodes).}
 \label{fig:model2result4-40}
\end{figure}

The values of the generalized clustering coefficients of the whole network are plotted in Figure \ref{fig:model2result40-4}, Figure \ref{fig:model2result8-8} and Figure \ref{fig:model2result4-40} respectively.
\begin{table*}[htbp]
\begin{center}
\caption{Values of the generalized clustering coefficients of the whole network near weight 0 (Model 2: upper = 40 nodes, lower = 4 nodes).}
\label{model2table40-4}
\begingroup
\renewcommand{\arraystretch}{1.3}
\scriptsize
\begin{tabular}{|l|c|c|c|c|} \hline
(Existing) & $\epsilon = 0$ & $\epsilon = 1^{-13}$ & $\epsilon = 1^{-12}$ & $\epsilon = 1^{-11}$ \\\hline
$C_{ops,am}$	& 0.1900 (100.0\%) & 1 (526.3\%) & 1 (526.3\%) & 1 (526.3\%) \\
$C_{ops,gm}$	& 0.1900 (100.0\%) & 1 (526.3\%) & 1 (526.3\%) & 1 (526.3\%) \\
$C_{ops,min}$	& 0.1900 (100.0\%) & 1 (526.3\%) & 1 (526.3\%) & 1 (526.3\%) \\
$C_{ops,max}$	& 0.1900 (100.0\%) & 1 (526.3\%) & 1 (526.3\%) & 1 (526.3\%) \\\hline
(Proposed) & $\epsilon = 0$ & $\epsilon = 1^{-13}$ & $\epsilon = 1^{-12}$ & $\epsilon = 1^{-11}$ \\\hline
$C_{glo,mu}$	& 0.1900 (100.0\%) & 0.1900 (100.0\%) & 0.1900 (100.0\%) & 0.1900 (100.0\%) \\
$C_{glo,gm}$	& 0.1900 (100.0\%) & 0.1901 (100.1\%) & 0.1902(100.1\%)  & 0.1905 (100.3\%) \\
$C_{glo,min}$	& 0.1900 (100.0\%) & 0.1900 (100.0\%) & 0.1900 (100.0\%) & 0.1900 (100.0\%) \\
$C_{glo,hm}$	& 0.1900 (100.0\%) & 0.1900 (100.0\%) & 0.1900 (100.0\%) & 0.1900 (100.0\%) \\\hline
\end{tabular}
\endgroup
\end{center}
\end{table*}

\begin{table*}[htbp]
\begin{center}
\caption{Values of the generalized clustering coefficients of the whole network near weight 0 (Model 2: upper = 8 nodes, lower = 8 nodes).}
\label{model2table8-8}
\begingroup
\renewcommand{\arraystretch}{1.3}
\scriptsize
\begin{tabular}{|l|c|c|c|c|} \hline
(Existing) & $\epsilon = 0$ & $\epsilon = 1^{-13}$ & $\epsilon = 1^{-12}$ & $\epsilon = 1^{-11}$ \\\hline
$C_{ops,am}$	& 0.7895 (100.0\%) & 1 (126.6\%) & 1 (126.6\%) & 1 (126.6\%) \\
$C_{ops,gm}$	& 0.7895 (100.0\%) & 1 (126.6\%) & 1 (126.6\%) & 1 (126.6\%) \\
$C_{ops,min}$	& 0.7895 (100.0\%) & 1 (126.6\%) & 1 (126.6\%) & 1 (126.6\%) \\
$C_{ops,max}$	& 0.7895 (100.0\%) & 1 (126.6\%) & 1 (126.6\%) & 1 (126.6\%) \\\hline
(Proposed) & $\epsilon = 0$ & $\epsilon = 1^{-13}$ & $\epsilon = 1^{-12}$ & $\epsilon = 1^{-11}$ \\\hline
$C_{glo,mu}$	& 0.7895 (100.0\%) & 0.7895 (100.0\%) & 0.7895 (100.0\%) & 0.7895 (100.0\%) \\
$C_{glo,gm}$	& 0.7895 (100.0\%) & 0.7895 (100.0\%)  & 0.7895 (100.0\%)   & 0.7895 (100.0\%)   \\
$C_{glo,min}$	& 0.7895 (100.0\%) & 0.7895 (100.0\%) & 0.7895 (100.0\%) & 0.7895 (100.0\%) \\
$C_{glo,hm}$	& 0.7895 (100.0\%) & 0.7895 (100.0\%) & 0.7895 (100.0\%) & 0.7895 (100.0\%) \\\hline
\end{tabular}
\endgroup
\end{center}
\end{table*}

\begin{table*}[htbp]
\begin{center}
\caption{Values of the generalized clustering coefficients of the whole network near weight 0 (Model 2: upper = 4 nodes, lower = 40 nodes).}
\label{model2table4-40}
\begingroup
\renewcommand{\arraystretch}{1.3}
\scriptsize
\begin{tabular}{|l|c|c|c|c|} \hline
(Existing) & $\epsilon = 0$ & $\epsilon = 1^{-13}$ & $\epsilon = 1^{-12}$ & $\epsilon = 1^{-11}$ \\\hline
$C_{ops,am}$	& 0.9939 (100.0\%) & 1 (100.6\%) & 1 (100.6\%) & 1 (100.6\%) \\
$C_{ops,gm}$	& 0.9939 (100.0\%) & 1 (100.6\%) & 1 (100.6\%) & 1 (100.6\%) \\
$C_{ops,min}$	& 0.9939 (100.0\%) & 1 (100.6\%) & 1 (100.6\%) & 1 (100.6\%) \\
$C_{ops,max}$	& 0.9939 (100.0\%) & 1 (100.6\%) & 1 (100.6\%) & 1 (100.6\%) \\\hline
(Proposed) & $\epsilon = 0$ & $\epsilon = 1^{-13}$ & $\epsilon = 1^{-12}$ & $\epsilon = 1^{-11}$ \\\hline
$C_{glo,mu}$	& 0.9939 (100.0\%) & 0.9939 (100.0\%) & 0.9939 (100.0\%) & 0.9939 (100.0\%) \\
$C_{glo,gm}$	& 0.9939 (100.0\%) & 0.9939 (100.0\%)  & 0.9939 (100.0\%) & 0.9939 (100.0\%)   \\
$C_{glo,min}$	& 0.9939 (100.0\%) & 0.9939 (100.0\%) & 0.9939 (100.0\%) & 0.9939 (100.0\%) \\
$C_{glo,hm}$	& 0.9939 (100.0\%) & 0.9939 (100.0\%) & 0.9939 (100.0\%) & 0.9939 (100.0\%) \\\hline
\end{tabular}
\endgroup
\end{center}
\end{table*}

In model 2, it is preferable that the coefficient values of the whole network approach to 0.1900, 0.7895 and 0.9939 respectively when $\epsilon$ is reduced from 1 to 0. However, in the case of the generalizations by Opsahl et al. \cite{opsahl2009clustering}, the coefficient values did not approach to the preferable values. On the other hand, the values by the proposed coefficients approached to the preferable values.

\subsection{Real World Network Datasets}

In this subsection, we describe the results of the experiments on the real world network datasets.

\subsubsection{Experiment 1}
We listed the coefficient values of Experiment 1 in Table \ref{table:real1-1} (Freeman EIES network) and Table \ref{table:real1-2} (C. Elegans Neural network). 

\begin{table*}[htbp]
\begin{center}
\caption{Values of the existing and our generalizations of the global clustering coefficient.(Experiment 1: Freeman EIES network)}
\label{table:real1-1}
\begingroup
\renewcommand{\arraystretch}{1.3}
\scriptsize
\begin{tabular}{|l|c|c|c|c|} \hline
(Existing) & $0$ & $10^{-12}$ &  $10^{-9}$ & $10^{-6}$ \\ \hline
$C_{ops,am}$ & 0.7378 (100.0\%) & 1 (135.5\%) & 1 (135.5\%) & 1 (135.5\%) \\
$C_{ops,gm}$ & 0.7332 (100.0\%) & 1 (136.4\%) & 1 (136.4\%) & 1 (136.4\%) \\
$C_{ops,min}$ & 0.7250 (100.0\%) & 1 (137.9\%) & 1 (137.9\%) & 1 (137.9\%) \\
$C_{ops,max}$ & 0.7411 (100.0\%) & 1 (134.9\%) & 1 (134.9\%) & 1 (134.9\%)  \\\hline
(Proposed) & $0$ & $10^{-12}$ &  $10^{-9}$ & $10^{-6}$ \\ \hline
$C_{glo,mu}$ & 0.1325 (100.0\%) & 0.1325 (100.0\%) & 0.1325 (100.0\%) & 0.1325 (100.0\%) \\
$C_{glo,gm}$ & 0.2688 (100.0\%) & 0.2688 (100.0\%) & 0.2687 (99.9\%) & 0.2673 (99.4\%) \\
$C_{glo,min}$ & 0.4678 (100.0\%) & 0.4678 (100.0\%) & 0.4678 (100.0\%) & 0.4678 (100.0\%) \\
$C_{glo,hm}$ & 0.4045 (100.0\%) & 0.4045 (100.0\%) & 0.4045 (100.0\%) & 0.4045 (100.0\%) \\\hline
\end{tabular}
\endgroup
\end{center}
\end{table*}

\begin{table*}[htbp]
\begin{center}
\caption{Values of the existing and our generalizations of the global clustering coefficient.(Experiment 1: C. Elegans Neural network)}
\label{table:real1-2}
\begingroup
\renewcommand{\arraystretch}{1.3}
\scriptsize
\begin{tabular}{|l|c|c|c|c|} \hline
(Existing) & $0$ & $10^{-12}$ &  $10^{-9}$ & $10^{-6}$ \\ \hline
$C_{ops,am}$ & 0.2364 (100.0\%) & 1 (422.9\%) & 1 (422.9\%) & 1 (422.9\%) \\
$C_{ops,gm}$ & 0.2179 (100.0\%) & 1 (459.0\%) & 1 (459.0\%) & 1 (459.0\%) \\
$C_{ops,min}$ & 0.2003 (100.0\%) & 1 (499.1\%) & 1 (499.1\%) & 1 (499.1\%) \\
$C_{ops,max}$ & 0.2475 (100.0\%) & 1 (404.0\%) & 1 (404.0\%) & 1 (404.0\%)  \\\hline
(Proposed) & $0$ & $10^{-12}$ &  $10^{-9}$ & $10^{-6}$ \\ \hline
$C_{glo,mu}$ & 0.0099 (100.0\%) & 0.0099 (100.0\%) & 0.0099 (100.0\%) & 0.0099 (100.0\%) \\
$C_{glo,gm}$ & 0.0401 (100.0\%) & 0.0401 (100.0\%) & 0.0397 (99.0\%) & 0.0367 (91.6\%) \\
$C_{glo,min}$ & 0.0709 (100.0\%) & 0.0709 (100.0\%) & 0.0709 (100.0\%) & 0.0712 (100.5\%) \\
$C_{glo,hm}$ & 0.0594 (100.0\%) & 0.0594 (100.0\%) & 0.0594 (100.0\%) & 0.0600 (100.3\%) \\\hline
\end{tabular}
\endgroup
\end{center}
\end{table*}

The values of the generalizations by Opsahl et al. \cite{opsahl2009clustering} was significantly changed. In the case of C. Elegans Neural network, $C_{ops,am}$ changed about 423 \% from the original value. 
On the other hand, the values of our generalizations did not significantly changed.

\subsubsection{Experiment 2}
We listed the coefficient values of Experiment 2 in Table \ref{table:real2-1} (Freeman EIES network) and Table \ref{table:real2-2} (C. Elegans Neural network).

\begin{table*}[htbp]
\begin{center}
\caption{Values of the existing and our generalizations of the global clustering coefficient.(Experiment 2: Freeman EIES network)}
\label{table:real2-1}
\begingroup
\renewcommand{\arraystretch}{1.3}
\scriptsize
\begin{tabular}{|l|c|c|c|c|} \hline
(Existing) & $0$ & $10^{-12}$ &  $10^{-9}$ & $10^{-6}$ \\ \hline
$C_{ops,am}$ & 0.7378 (100.0\%) & 0.9172 (124.3\%) & 0.9141 (123.9\%) & 0.9146 (124.0\%) \\
$C_{ops,gm}$ & 0.7332 (100.0\%) & 0.9583 (130.7\%) & 0.9575 (130.6\%) & 0.9568 (130.5\%) \\
$C_{ops,min}$ & 0.7250 (100.0\%) & 0.9579 (132.1\%) & 0.9564 (131.9\%) & 0.9563 (132.0\%) \\
$C_{ops,max}$ & 0.7411 (100.0\%) & 0.9127 (123.1\%) & 0.9094 (122.7\%) & 0.9100 (122.8\%)  \\\hline
(Proposed) & $0$ & $10^{-12}$ &  $10^{-9}$ & $10^{-6}$ \\ \hline
$C_{glo,mu}$ & 0.1325 (100.0\%) & 0.1325 (100.0\%) & 0.1325 (100.0\%) & 0.1325 (100.0\%) \\
$C_{glo,gm}$ & 0.2688 (100.0\%) & 0.2688 (100.0\%) & 0.2687 (99.9\%) & 0.2675 (99.5\%) \\
$C_{glo,min}$ & 0.4678 (100.0\%) & 0.4678 (100.0\%) & 0.4678 (100.0\%) & 0.4678 (100.0\%) \\
$C_{glo,hm}$ & 0.4045 (100.0\%) & 0.4045 (100.0\%) & 0.4045 (100.0\%) & 0.4045 (100.0\%) \\\hline
\end{tabular}
\endgroup
\end{center}
\end{table*}

\begin{table*}[htbp]
\begin{center}
\caption{Values of the existing and our generalizations of the global clustering coefficient.(Experiment 2: C. Elegans Neural network)}
\label{table:real2-2}
\begingroup
\renewcommand{\arraystretch}{1.3}
\scriptsize
\begin{tabular}{|l|c|c|c|c|} \hline
(Existing) & $0$ & $10^{-12}$ &  $10^{-9}$ & $10^{-6}$ \\ \hline
$C_{ops,am}$ & 0.2364 (100.0\%) & 0.8551 (361.7\%) & 0.8554 (361.8\%) & 0.8555 (361.8\%) \\
$C_{ops,gm}$ & 0.2179 (100.0\%) & 0.8751 (401.7\%) & 0.8779 (402.9\%) & 0.8760 (402.0\%) \\
$C_{ops,min}$ & 0.2003 (100.0\%) & 0.8723 (435.4\%) & 0.8746 (436.5\%) & 0.8739 (436.2\%) \\
$C_{ops,max}$ & 0.2475 (100.0\%) & 0.8550 (345.4\%) & 0.8552 (345.5\%) & 0.8553 (345.5\%)  \\\hline
(Proposed) & $0$ & $10^{-12}$ &  $10^{-9}$ & $10^{-6}$ \\ \hline
$C_{glo,mu}$ & 0.0099 (100.0\%) & 0.0099 (100.0\%) & 0.0099 (100.0\%) & 0.0099 (100.0\%) \\
$C_{glo,gm}$ & 0.0401 (100.0\%) & 0.0401 (99.9\%) & 0.0398 (99.2\%) & 0.0371 (92.6\%) \\
$C_{glo,min}$ & 0.0709 (100.0\%) & 0.0709 (100.0\%) & 0.0709 (100.0\%) & 0.0710 (100.2\%) \\
$C_{glo,hm}$ & 0.0594 (100.0\%) & 0.0594 (100.0\%) & 0.0594 (100.0\%) & 0.0595 (100.2\%) \\\hline
\end{tabular}
\endgroup
\end{center}
\end{table*}

The values of the generalizations by Opsahl et al. \cite{opsahl2009clustering} was significantly changed. In the case of C. Elegans Neural network, $C_{ops,am}$ changed about 361 \% from the original value. On the other hand, the values of our generalizations did not significantly changed.

These results suggest that the larger network, the larger degree of error.

\section{Conclusions}
\label{section6}

In this paper, we focused on the continuity of the generalized clustering coefficient, and proposed continuous generalized clustering coefficients for weighted and directed networks.
All of the existing generalizations for weighted and directed networks and generalizations of the global clustering coefficient are discontinuous.
We examined the robustness against the edge weight errors of our continuous generalized clustering coefficients by experiments on artificial networks and real world networks. 
In the case of the Experiment 2 of the C. Elegans Neural network, though the value of the one existing generalization (method: minimum) was changed about 436 \%, the value of a proposed one (method: minimum) was only changed 0.2 \%.

In future work, we need to show the case that the continuity of generalized clustering coefficients has a more concrete impact on practical research. 

\section*{Acknowledgement}

The authors wish to thank Tore Opsahl and Pietro Panzarasa for providing the datasets.

\bibliographystyle{plain}
\bibliography{else}

\begin{thebibliography}{10}

\bibitem{abdallah2009new}
S.~Abdallah.
\newblock {A New Methodology for Generalizing Unweighted Network Measures}.
\newblock {\em Arxiv preprint arXiv:0905.4169}, 2009.

\bibitem{ahnert2007ensemble}
SE~Ahnert, D.~Garlaschelli, TMA Fink, and G.~Caldarelli.
\newblock {Ensemble approach to the analysis of weighted networks}.
\newblock {\em Physical Review E}, 76(1):16101, 2007.

\bibitem{barrat2004architecture}
A.~Barrat, M.~Barth{\'e}lemy, R.~Pastor-Satorras, and A.~Vespignani.
\newblock {The architecture of complex weighted networks}.
\newblock {\em Proceedings of the National Academy of Sciences}, 101(11):3747,
  2004.

\bibitem{fagiolo2007clustering}
G.~Fagiolo.
\newblock {Clustering in complex directed networks}.
\newblock {\em Physical Review E}, 76(2):26107, 2007.

\bibitem{freeman1979networkers}
S.C. Freeman and L.C. Freeman.
\newblock {\em {The networkers network: A study of the impact of a new
  communications medium on sociometric structure}}.
\newblock School of Social Sciences Univ. of Calif., 1979.

\bibitem{grindrod2002range}
P.~Grindrod.
\newblock {Range-dependent random graphs and their application to modeling
  large small-world proteome datasets}.
\newblock {\em Physical Review E}, 66(6):66702, 2002.

\bibitem{holme2007korean}
P.~Holme, S.~Min~Park, B.J. Kim, and C.R. Edling.
\newblock {Korean university life in a network perspective: Dynamics of a large
  affiliation network}.
\newblock {\em Physica A: Statistical and Theoretical Physics}, 373:821--830,
  2007.

\bibitem{ide2007limit}
Y.~Ide, N.~Konno, and N.~Masuda.
\newblock {Limit theorems for some statistics of a generalized threshold}.
\newblock {\em RIMS Kokyuroku}, 1551:81--86, 2007.

\bibitem{kalna2006clustering}
G.~Kalna and D.J. Higham.
\newblock {Clustering coefficients for weighted networks}.
\newblock {\em University of Strathclyde Mathematics Research Report}, 3, 2006.

\bibitem{konno_ide200805}
N.~Konnno and Y.~Ide.
\newblock {\em Introduction to Complex Networks.}
\newblock Kodansha, 5 2008.
\newblock Japanese.

\bibitem{latora2003economic}
V.~Latora and M.~Marchiori.
\newblock {Economic small-world behavior in weighted networks}.
\newblock {\em The European Physical Journal B-Condensed Matter and Complex
  Systems}, 32(2):249--263, 2003.

\bibitem{lopez2004applying}
L.~Lopez-Fernandez, G.~Robles, and J.M. Gonzalez-Barahona.
\newblock {Applying social network analysis to the information in cvs
  repositories}.
\newblock In {\em International Workshop on Mining Software Repositories}.
  Citeseer, 2004.

\bibitem{newman2002random}
M.E.J. Newman, D.J. Watts, and S.H. Strogatz.
\newblock {Random graph models of social networks}.
\newblock {\em Proceedings of the National Academy of Sciences}, 99(Suppl
  1):2566, 2002.

\bibitem{onnela2005intensity}
J.P. Onnela, J.~Saram{\\"a}ki, J.~Kert{\'e}sz, and K.~Kaski.
\newblock {Intensity and coherence of motifs in weighted complex networks}.
\newblock {\em Physical Review E}, 71(6):65103, 2005.

\bibitem{opsahl2009clustering}
T.~Opsahl and P.~Panzarasa.
\newblock {Clustering in weighted networks}.
\newblock {\em Social networks}, 31(2):155--163, 2009.

\bibitem{schank2005approximating}
T.~Schank and D.~Wagner.
\newblock {Approximating clustering coefficient and transitivity}.
\newblock {\em Journal of Graph Algorithms and Applications}, 9(2):265--275,
  2005.

\bibitem{suzuki2009fukuzatsu}
T.~Suzuki.
\newblock {Analysis for Directed and Weighted Complex Networks on the Basis of
  Information Flow}.
\newblock {\em IPSJ Transactions on Mathematical Modeling and its
  Applications}, 2(1):70--79, 2009.
\newblock Japanese with English summary.

\bibitem{wasserman1994social}
S.~Wasserman and K.~Faust.
\newblock {\em {Social network analysis: Methods and applications}}.
\newblock Cambridge Univ Pr, 1994.

\bibitem{watts1998collective}
D.J. Watts and S.H. Strogatz.
\newblock {Collective dynamics of esmall-worldfnetworks}.
\newblock {\em Nature}, 393(6684):440--442, 1998.

\bibitem{zhang2005general}
B.~Zhang and S.~Horvath.
\newblock {A general framework for weighted gene co-expression network
  analysis}.
\newblock {\em Statistical Applications in Genetics and Molecular Biology},
  4(1):1128, 2005.

\end{thebibliography}

\end{document}